\documentclass[11pt,a4paper]{article}
\pdfoutput=1
\usepackage{jcappubm}
\usepackage[british]{babel}
\usepackage{latexsym,amsmath,amssymb}
\usepackage{graphicx}
\usepackage{booktabs}

\newcommand{\eq}[1]{(\ref{#1})}
\def \cshat{\hat{c_X}^2}
\def \ch{{\cal H}}

    \newcommand{\be}{\begin{eqnarray}}
    \newcommand{\ee}{\end{eqnarray}}
    \newcommand{\bea}{\begin{eqnarray}}
    \newcommand{\eea}{\end{eqnarray}}
    \newcommand{\ba}{\begin{array}}
    \newcommand{\ea}{\end{array}}
    \newcommand{\nn}{\nonumber}

    \newcommand{\Mpc}{{\rm Mpc}}
    \newcommand{\km}{{\rm km}}

    \renewcommand{\(}{\left(}
    \renewcommand{\)}{\right)}
    \renewcommand{\[}{\left[}
    \renewcommand{\]}{\right]}
    \newcommand{\la}{\langle}
    \newcommand{\ra}{\rangle}

    \newcommand{\mn}{{\mu\nu}}
    \newcommand{\ab}{{\alpha\beta}}
    \newcommand{\vf}{\varphi}
    \newcommand{\HH}{\mathcal{H}}
    \newcommand{\dd}{\partial}

    \newcommand{\dln}{\dd\!\ln}

    \newcommand{\Ve}[1]{\mathbf{#1}}
    \newcommand{\Vx}{\Ve{x}}
    
    \newcommand{\Vk}{\Ve{k}}
    \newcommand{\Vq}{\Ve{q}}
    
    \newcommand{\Vnhat}{\hat{\Ve{n}}}
    
    \renewcommand{\H}{{\cal H}}

    \renewcommand{\P}{{\cal P}}    
    \newcommand{\veps}{\varepsilon}

    \newcommand{\br}{{\bar\rho}}
    \newcommand{\bp}{{\bar p}}
    \newcommand{\der}{{\delta\rho}}
    \newcommand{\dep}{{\delta{}p}}
    \newcommand{\deq}{{\delta{}q}}

    \newcommand{\depi}{{\delta\pi}}

\definecolor{MyBlue}{rgb}{0,0,0.8}

\definecolor{darkgreen}{rgb}{0.1,0.75,0.1}

\definecolor{MyRed}{rgb}{0.9,0.12,0.1}

\author[a,b,c]{Guillermo Ballesteros,}
\author[d]{Lukas Hollenstein,}
\author[d]{Rajeev Kumar Jain}
\author[d]{and Martin Kunz}
\affiliation[a]{Museo Storico della Fisica e Centro Studi e Ricerche ``Enrico Fermi'', Piazza del Viminale 1, I-00184 Rome, Italy}
\affiliation[b]{Dipartimento di Fisica ``G.\ Galilei'', Universit\`a degli Studi di Padova, Via Marzolo 8, I-35131 Padua, Italy}
\affiliation[c]{INFN, Sezione di Padova, Via Marzolo 8, I-35131 Padua, Italy}
\affiliation[d]{D\'epartement de Physique Th\'eorique and Center for Astroparticle Physics, Universit\'e de Gen\`eve, Quai E.\ Ansermet 24, CH-1211 Gen\`eve 4, Switzerland}

\emailAdd{guillermo.ballesteros@pd.infn.it}
\emailAdd{lukas.hollenstein@unige.ch}
\emailAdd{rajeev.jain@unige.ch}
\emailAdd{martin.kunz@unige.ch}

\begin{document}

\title{
Nonlinear cosmological consistency relations and effective matter stresses}

\abstract{
We propose a fully nonlinear framework to construct consistency relations for testing generic cosmological scenarios using the evolution of large scale structure. It is based on the covariant approach in combination with a frame that is purely given by the metric, the normal frame. As an example, we apply this framework to the $\Lambda$CDM model, by extending the usual first order conditions on the metric potentials to second order, where the two potentials start to differ from each other. We argue that working in the normal frame is not only a practical choice but also helps with the physical interpretation of nonlinear dynamics. In this frame, effective pressures and anisotropic stresses appear at second order in perturbation theory, even for ``pressureless'' dust. We quantify their effect and compare them, for illustration, to the pressure of a generic clustering dark energy fluid and the anisotropic stress in the DGP model. Besides, we also discuss the effect of a mismatch of the potentials on the determination of galaxy bias.
}


\arxivnumber{1112.4837}

\maketitle

\section{Introduction}  \label{sec:intro}

Current cosmological observations provide evidence for a recent onset of a phase of accelerated expansion of the Universe \cite{Riess:1998cb, Perlmutter:1998np, Komatsu:2010fb,Suzuki:2011hu,Sherwin:2011gv}. Under the assumption of large-scale homogeneity and isotropy, the observations can only be described in the framework of General Relativity (GR) by invoking some form of Dark Energy (DE) with negative pressure. Apart from the Standard Model (SM) of particle physics, the preferred model of the content of the Universe also assumes the presence of Cold Dark Matter (CDM) and DE in the form of a cosmological constant ($\Lambda$) and is therefore named $\Lambda$CDM. The dynamics of the Universe is based on the Einstein equations and on the assumption that the average geometry is described by a Friedmann-Lema\^itre-Robertson-Walker (FLRW) metric. The formation of structures is then modelled using cosmological perturbation theory \cite{Bardeen:1980kt, Kodama:1985bj, Mukhanov:1990me, Durrer:1993db, Ma:1995ey, Bernardeau:2001qr} which describes the evolution of small fluctuations in the energy-momentum content and the metric.

The $\Lambda$CDM model comes with a minimal set of parameters that are increasingly well constrained by the data and, to date, no significant departures from its predictions on cosmological scales have been found \cite{Komatsu:2010fb, Bean:2010zq, Zhao:2010dz}. However, the difficulty in explaining the required tiny value of $\Lambda$ has motivated many alternative scenarios with dynamical DE fields or modifications to GR aka Modified Gravity (MG)~\cite{Wetterich:1987fm, Ratra:1987rm, ArmendarizPicon:1999rj, Dvali:2000hr, Deffayet:2001pu, Copeland:2006wr, Durrer:2008in, Tsujikawa:2010zza, Clifton:2011jh, Capozziello:2011et,Kunz:2012aw}. Given the existing tight constraints, it is essential  to understand very accurately what the theoretical predictions are and how to interpret the observables, in order to maximise the benefits of the future data. A possible falsification of $\Lambda$CDM would then be the first step towards a complete understanding of the dark energy. In practice, we can choose between comparing $\Lambda$CDM to a large representative number of alternative scenarios, parameterise departures from it in a sufficiently flexible (but still economical) way \cite{Caldwell:2007cw, Amendola:2007rr, Hu:2007pj, Song:2008vm, Bertschinger:2008zb, Song:2010rm, Pogosian:2010tj, Daniel:2010yt, Song:2010fg, Hojjati:2011ix, Zhao:2011te, Dossett:2011zp, Dossett:2011tn, Baker:2011jy, Zuntz:2011aq}, or identify key relations whose breakdown would hint towards a falsification of the standard paradigm \cite{Huterer:2006mva, Song:2008vm, Song:2008xd}. All three strategies are useful in their own right: the third option is initially most powerful for falsifying or confirming $\Lambda$CDM, while the second can be helpful as a guidance for the first, more traditional option. In any case, exquisite precision in the model predictions and understanding of the observables are needed to make decisive statements.

In this work we propose and start to explore a program for constructing fully nonlinear consistency relations between geometry and matter content, not assuming GR but allowing for generic modifications of the field equations. This can be seen as a (nonlinear) step beyond the popular modified growth parameterisations \cite{Caldwell:2007cw, Amendola:2007rr, Hu:2007pj, Song:2008vm, Bertschinger:2008zb, Song:2010rm, Pogosian:2010tj, Daniel:2010yt, Song:2010fg, Hojjati:2011ix, Zhao:2011te, Dossett:2011zp, Dossett:2011tn, Baker:2011jy, Zuntz:2011aq}. Such relations can either be implemented as consistency checks to falsify $\Lambda$CDM (or some other paradigm), or can be parameterised to explore a possible departure from $\Lambda$CDM. To construct the consistency relations we make use of the fact that at late times matter (baryonic and dark) is well described by a pressureless perfect fluid (dust) on cosmological scales. The absence of isotropic and anisotropic pressures in the matter rest frame directly implies two consistency relations between the geometry and the physics that gives rise to the late time accelerated expansion. However, we point out some important subtleties that arise due to the choice of frame and that can become relevant on nonlinear scales.

We first formulate the consistency relations in the exact nonlinear theory employing the covariant approach \cite{Ellis:1971pg, Ellis:1989jt, Bruni:1992dg, Ehlers:1993gf, Maartens:1998xg,Clarkson:2010uz}. Then, we discuss how the conditions come about after matter-radiation equality in first and second order cosmological perturbation theory, using the longitudinal (or conformal Newtonian) gauge. At first order, the resulting conditions are well known. They simply tell us that the two metric potentials are equal, $\phi_1=\psi_1$, and their evolution is governed by a simple second order differential equation that can be solved exactly: the Bardeen equation \cite{Bardeen:1980kt}. At second order, the total measured pressure and anisotropic stress depend on the observer frame that is used to define these quantities, and this is where the subtleties come into play \cite{Hwang:2005hd}. The pressure measured in the frame comoving with the matter is zero by definition (in analogy to the Lagrangian picture in the Newtonian approximation, see for instance \cite{Hwang:2006iw,Villa:2011vt}). However, in a non-comoving frame, effective pressures and anisotropic stresses are induced by the relative velocity field, even if the source is a pressureless perfect fluid \cite{Hwang:2005hd}. As we shall discuss in detail, there are good reasons to work in a specific non-comoving frame, called the \emph{normal frame}. This frame is orthogonal to the surfaces of constant time and is given purely by the geometry, i.e.\ the (perturbed) metric. An added advantage of this framework is that the density and pressure of matter (and of any other component) in the normal frame are then immediately related to the \emph{geometrical pressures} that can be read off from the Einstein tensor. This means that projecting on the normal frame lets us cleanly separate the geometrical fluctuations from those of the stress-energy content. Consequently, the consistency relations in the normal frame take a particularly simple form. The change of frame from comoving to non-comoving induces effective matter pressures that we quantify. We find them to be small in comparison to typical non-standard sources of pressure and anisotropic stress in models beyond $\Lambda$CDM.

The outline of the paper is as follows. First, in section \ref{sec:covariant}, we review the effects of changing the observer frame from comoving to non-comoving, and discuss why the normal frame is particularly useful. In section \ref{sec:metric_linear} we state the consistency relations in the exact nonlinear case, discuss the connection between the covariant and the metric perturbations approach and review the first order consistency relations. In section \ref{sec:nonlinear} we derive the second order consistency relations and compute the effective matter pressure and anisotropic stress. Section \ref{sec:discussion} is devoted to the discussion of a number of typical cases where the effective matter stresses could potentially lead to wrong physical interpretations  of the data. We consider the cases of a general clustering DE fluid and the DGP (Dvali-Gabadadze-Porrati \cite{Dvali:2000hr}) model as a prototypical example of MG, because for both their respective intrinsic pressure and/or anisotropic stress perturbations could, in principle, be confused with the contributions from matter. Further, we discuss the impact on the determination of the galaxy bias from weak lensing and galaxy clustering. We summarise our work and conclude in section \ref{sec:conclusions}. We include three appendices. In appendix \ref{app:conventions} we state our conventions on notation and units. Appendix \ref{app:covariant} briefly reviews the covariant approach and appendix \ref{app:metric} gives more details on the metric perturbation calculations.

\section{Covariant approach and change of frame} \label{sec:covariant}

The covariant approach is very useful to formally describe the full nonlinear dynamics in terms of physical variables that are covariantly defined as projections with respect to a family of observers, i.e.\ a frame. In this section, we first discuss the notion of observer frame in the covariant approach and its relation to the choice of coordinates and gauges in cosmological perturbation theory. Then, we focus on the effect of changing the frame in the case of pure pressureless dust, which forms the basis for stating the fully nonlinear consistency relations in the next section.

\subsection{Frames and gauges}  \label{sec:framesgauges}

The covariant approach is based on the choice of a unit time-like 4-velocity that defines a 1+3 splitting of space-time and a unique decomposition of tensorial quantities into irreducible scalars, and projected vectors and tensors. Such a velocity field defines an \emph{observer frame} and is generically denoted by $u^\mu$ (with $u_\alpha u^\alpha=-1$). The general decomposition of an energy-momentum tensor
\be \label{eq:Tdecomp}
T^\mn = \rho u^\mu u^\nu + p \P^\mn + 2q^{(\mu} u^{\nu)} + \pi^\mn
\ee
yields the dynamical quantities (or \emph{fluid variables}): energy density $\rho$, pressure $p$, energy flux $q^\mu$ and anisotropic stress $\pi^\mn$ as measured by a family of observers with 4-velocity $u^\mu$. The tensor $\P^\mn\equiv g^\mn+u^\mu u^\nu$ is a projector on hypersurfaces orthogonal to $u^\mu$ and we follow the convention that indices in round brackets are symmetrised. For more details on the covariant formalism and the relevant definitions we refer the reader to appendix \ref{app:covariant_defs}.

Let us remark that an observer frame is not a choice of \emph{gauge} (or coordinates). The dynamical quantities in a given frame can in general be described in any gauge. However, the coordinates in which a specific frame takes its simplest form define a gauge which is intrinsically related to the choice of frame. For example, the comoving frame of matter (defined by the matter 4-velocity) takes its trivial form $u_m^\mu=a^{-1}\delta^\mu_0$ in the synchronous comoving gauge where the matter peculiar velocity vanishes. In this sense, a choice of frame prefers (but does not demand) the specific gauge in which the frame is trivial.

Choosing the comoving frame to do perturbation theory in curved space is analogous to using the Lagrangian picture in the Newtonian approximation, where the observer moves with the fluid flow. On the other hand, non-comoving frames can be seen as the analogues to the Eulerian framework in the Newtonian limit, where the observer and the coordinates are independent of the fluid and the properties of the fluid change at fixed coordinates in 3-space as time flows. However, in curved space-time there are infinitely many frames that are not comoving with, say, the matter. We can always translate one choice of frame into another but, for practical purposes, the question is how to choose a useful one. A hint can be found from the fact that non-comoving frames that see neither shear nor vorticity, called quasi-Newtonian frames, are those that provide the most direct connection between general relativistic and Newtonian cosmology, as argued e.g.\ in  \cite{Maartens:1998qw,vanElst:1998kb}.

For any given metric, the lapse and shift, $N$ and $N^i$, of the Arnowitt-Deser-Misner (ADM) 1+3 formalism can be used to define a covariant vector field normal to the spatial hypersurfaces, $n_\mu=-N\delta_\mu^0$, $n^\mu=N^{-1}(1,\,-N^i)$ \cite{Arnowitt:1962hi}. In a perturbed FLRW metric this simply becomes
\be
n_\mu \propto \frac{\partial\eta}{\partial x^\mu}
\ee
where $n_\alpha n^\alpha=-1$ and $\eta$ is the conformal time. We will refer to the vector $n^\mu$ as the \emph{normal frame}. Clearly, this frame is completely defined by the metric. The coordinates in which the normal frame has vanishing spatial components, $n^i=0=n_i$, are those of the longitudinal gauge. Moreover, the shear and the vorticity seen by $n^\mu$ vanish in this gauge, which means that the normal frame is quasi-Newtonian. The zero-shear condition for $n^\mu$ can in fact be used to define the longitudinal gauge at first as well as at second order in perturbations, see e.g.\ \cite{Malik:2008im}. In the appendix \ref{app:metric} we provide details on the properties of the normal frame and its kinematic quantities in the metric perturbation approach.

The normal frame $n^\mu$ is a non-comoving frame that is useful as a computational and conceptual tool for the following reasons. The evolution equations for matter in second order relativistic perturbation theory in the normal frame coincide with the Newtonian hydrodynamic equations in the Eulerian frame, see e.g.\ \cite{Hwang:2006iw}. Furthermore, the longitudinal gauge variables match the output of N-body simulations, except for the density perturbation that needs a gauge-correction on large (linear) scales where the Newtonian approximation starts to break down \cite{Chisari:2011iq, Green:2011wc}.  Finally, we will show in section \ref{sec:secconsistency} that the contributions of the stress-energy to the source terms of the evolution equations of the second order metric potentials are directly related to the pressure and anisotropic stress in the normal frame. This provides a nice physical interpretation of the normal frame quantities.

\subsection{Change of frame}  \label{sec:framechange}

Fluid variables computed in a given observer frame can be translated into those seen in another frame fully nonlinearly, see e.g.\ \cite{Maartens:1998xg, Clarkson:2010uz}. The transformations are given by the \emph{relative velocity} between the two frames, $v^\mu$. For the fluid variables all transformations are second order in the relative velocity, except for the energy flux. 

Here we focus on the transformation from the (comoving) rest frame of matter, denoted by $u_m^\mu$, to the (non-comoving) normal frame $n^\mu$. The general case can be found in appendix \ref{app:framechange}. By definition, the rest frame (also called energy frame) is the one in which the energy-flux vanishes \cite{Bruni:1992dg}. We denote the fluid variables measured in the rest frame with an asterisk and therefore $q_m^{*\mu}\equiv 0$. The rest frame velocity decomposes with respect to the normal frame as
\be 
u_m^\mu = \gamma_m \( n^\mu + v_m^\mu \)
\label{eq:relveldef}
\ee 
where $\gamma_m\equiv (1-v_m^2)^{-1/2}$ (with $v_m^2 \equiv v_{m\,\alpha} v_m^\alpha$) is the local Lorentz factor of the transformation, and the relative velocity $v_m^\mu$ is orthogonal to $n^\mu$ such that $n_\alpha v_m^\alpha=0$. The fluid variables as measured by the normal frame $n^\mu$ are related to the rest frame variables via
\be 	\label{trns1}
\rho_m &=& \gamma_m^2\rho_m^*
\\
p_m &=& \frac{1}{3}\gamma_m^2 v_m^2\rho_m^* = \frac{1}{3}v_m^2\rho_m
\label{eq:frametransform_pm}
\\
q_m^\mu &=& \rho_m^*v_m^\mu + \gamma_m^2v_m^2\rho_m^*v_m^\mu = \rho_m v_m^\mu
\\
\pi_m^\mn &=& \gamma_m^2\rho_m^*v_m^{\la\mu}v_m^{\nu\ra} = \rho_m v_m^{\la\mu}v_m^{\nu\ra}
\phantom{\frac{1}{3}}  \label{eq:frametransform_pim}
\ee
where we used the transformation of the energy density \eq{trns1} to express the effective pressure, energy flux and anisotropic stress in terms of the energy density in the normal frame and the relative velocity. The non-vanishing stresses present in the normal frame, $p_m$ and $\pi_m^\mn$, are therefore related to the motion of the matter particles. It is also instructive to realise that the pressure corresponds to the difference in energy density measured in the two frames
\be \label{eq:rhorhostar}
p_m = \frac{1}{3}\(\rho_m - \rho_m^*\)
\,,\qquad
\pi_m^\mn = \(\rho_m - \rho_m^*\) \frac{ v_m^{\la\mu}v_m^{\nu\ra} }{ v_m^2 }
\ee
and $\rho_m-\rho_m^*=v_m^2 \rho_m$. In other words: the trace of the energy-momentum tensor, $T^\alpha_\alpha=\rho-3p$ is invariant under the change of frame.

To get some more intuition, let us see how the non-comoving density, $\rho_m$, differs from the comoving density $\rho_m^*$. The mean pressure is related to the relative difference of the mean densities, defined as
\be
\nu \equiv \frac{\la\rho_m(x)\ra - \la\rho^*_m(x)\ra}{\la\rho_m(x)\ra}
= \frac{\la 3p_m(x)\ra}{\la\rho_m(x)\ra} \, . \label{eq:nu}
\ee
In second order perturbation theory, this turns out to be the dispersion of the relative velocity, see section \ref{sec:effectivepress}.
In terms of the two-point function $\xi(r) \equiv \la\rho_m(x)\,\rho_m(x+r)\ra / \la\rho_m(x)\ra^2 - 1$ we find that
the difference between the comoving and the non-comoving correlation function is given by
\be \label{eq:fullcorrelfunc}
\xi^*(r) - \xi(r) = \(2\nu-\nu^2\)\(\xi^*(r) +1\) 
+ \frac{\la 3p_m(x)\,3p_m(x+r)\ra -2\la\rho_m(x)\,3p_m(x+r)\ra}{\la\rho_m(x)\ra^2}
\ee
which nicely shows that this difference is mainly due to the velocity dispersion with corrections involving the pressure auto-correlation and pressure-density cross-correlation. Without further investigations we cannot say which is the precise correlation function actually observed. But since we do observe effects from peculiar velocities, like redshift space distortions, it appears unlikely that observations are not sensitive to the effective pressure and anisotropic stress.

To summarise, the fluid variables measured in the two types of frames coincide at first order in perturbations (except for the energy flux that is linked to the relative velocity). However, at second and higher order there is an intrinsic difference as effective stresses arise in non-comoving frames. We will quantify $p_m$ and $\pi_m^\mn$ in second order perturbation theory in section \ref{sec:effectivepress}. As we will see in the next section, the presence of the matter pressure in non-comoving frames translates into new terms in the nonlinear consistency relations between geometry and matter. For further insight into the role of matter pressure in non-comoving frames, see also \cite{Hwang:2005hd}.

\section{Exact consistency relations and linear theory}  \label{sec:metric_linear}

The currently most promising approach to learn about the physical nature of dark energy requires not just an accurate determination of its equation of state, but also of its pressure perturbation and anisotropic stress. In a $\Lambda$CDM universe, the pressure and the anisotropic stress vanish in the matter rest frame (apart from a small contribution from relativistic particles that is negligible at late times). Any deviation from zero would therefore be normally interpreted as a sign of non trivial dark energy dynamics. Let us now see how this consistency check comes about in the covariant approach and how its precise statement actually depends on the observer frame. Then we will make the connection to the metric perturbations approach and discuss the consistency relations in linear theory.

\subsection{Fully nonlinear $\Lambda$CDM consistency}  \label{sec:fullconsistency}

We start with a minimal set of assumptions before restricting the discussion to the $\Lambda$CDM scenario. We assume that matter follows the geodesics of a metric $g_\mn$ that gives rise to an Einstein tensor $G^\mn$. Matter refers here to baryonic matter, cold dark matter, radiation and neutrinos; and it is described by the total (matter) energy-momentum tensor $T^\mn$. Then,
\be
X^\mn \equiv M_P^2 G^\mn - T^\mn
\label{eq:Xdef}
\ee
is interpreted as the effective energy-momentum tensor of the dark physics causing the late time accelerated expansion. $X^\mn$ can represent a cosmological constant, any additional dark energy fields, modifications to GR, variations of Newton's constant or effective back-reaction effects. We refer to $X^\mn$ collectively as the \emph{dark energy}. The only assumption so far is that there is (effectively) a single metric that describes space-time and gravity. If we also assume that the theory is free of torsion, the Bianchi identities of the Einstein tensor read $\nabla_\alpha G^{\alpha\mu}=0$ and therefore yield $\nabla_\alpha(T^{\alpha\mu}+X^{\alpha\mu})=0$. Only under the further assumption that $T^\mn$ is covariantly conserved on its own we would also be allowed to write that $\nabla_\alpha X^{\alpha\mu}=0$, but we do not need to make this restriction here.

The path to get the full nonlinear consistency equations is to decompose \eqref{eq:Xdef} just like the energy-momentum tensor in \eqref{eq:Tdecomp}. In order to learn about the dark energy, we could then project it using its own 4-velocity  which is, though, unknown. Conversely, if we were to read off the dark energy properties in the matter rest frame, we would decompose \eqref{eq:Xdef} with respect to $u_m^\mu$ and deduce
\be
p_X^* = p_G^* \,, \qquad
\pi_X^{*\mn} = \pi_G^{*\mn} \,.
\ee
However, the \emph{geometric pressure and anisotropic stress} measured in the matter rest frame, $p_G^*$ and $\pi_G^{*\mn}$, are nonlinear combinations of matter and geometry variables that are hard to disentangle. Projecting on the normal frame will separate geometry and matter neatly. Therefore, we propose to decompose \eqref{eq:Xdef} in the normal frame $n^\mu$ and apply the transformations (\ref{eq:frametransform_pm}) and (\ref{eq:frametransform_pim}) to find
\be \label{eq:fullconsistency}
p_X = p_G - \frac{1}{3}v_m^2\rho_m \,, \qquad
\pi_X^\mn = \pi_G^\mn - \rho_m v_m^{\la\mu}v_m^{\nu\ra} \,.
\ee
Since the normal vector $n^\mu$ is defined exclusively by the metric, the geometric pressure and anisotropic stress measured in the normal frame, $p_G$ and $\pi_G^\mn$, are purely given in terms of the geometry, i.e.\ the metric potentials and the scale factor (for FLRW). Therefore, these quantities can in principle be reconstructed from purely geometrical observations such as Weak Lensing (WL), the Integrated Sachs-Wolfe (ISW) effect, the motion of test `particles' like galaxies, and the background expansion measurements. The contributions to the dark energy variables from the matter pressure and anisotropic stress can be reconstructed from observations of clustering, peculiar velocities and growth of structure. The exact nonlinear $\Lambda$CDM consistency relations are now simply deduced by requiring $p_X=-M_P^2\Lambda$ and $\pi^\mn_X=0$:
\be \label{eq:fullconsistencyLCDM}
-M_P^2 \Lambda = \bigg(p_G - \frac{1}{3}v_m^2\rho_m\bigg)_{\Lambda{\rm CDM}} \,, \qquad
0 = \bigg(\pi_G^\mn - \rho_m v_m^{\la\mu}v_m^{\nu\ra}\bigg)_{\Lambda{\rm CDM}}
\ee
Any breakdown of these relations would falsify the $\Lambda$CDM paradigm, and the dark pressure and anisotropic stress deduced from \eqref{eq:fullconsistency} would give information of the physics beyond $\Lambda$CDM. This is one of the main results of this paper: the velocity contribution to these consistency relations is absent in linear perturbation theory. The commonly used $\Lambda$CDM consistency relations at first order are therefore broken by this contribution.

\subsection{Comparison with modified growth approaches}
Let us now briefly connect this framework to the modified growth parameterisations of \cite{Caldwell:2007cw, Amendola:2007rr, Hu:2007pj, Song:2008vm, Bertschinger:2008zb, Song:2010rm, Pogosian:2010tj, Daniel:2010yt, Song:2010fg, Hojjati:2011ix, Zhao:2011te, Dossett:2011zp, Dossett:2011tn, Baker:2011jy, Zuntz:2011aq}. These are sometimes also referred to as Parameterised Post-Friedmannian approaches. First of all, notice that these formalisms are usually restricted to linear scalar perturbations, while the relations \eqref{eq:fullconsistency} and the $\Lambda$CDM consistency conditions \eqref{eq:fullconsistencyLCDM} are fully general, nonlinear and are not restricted to scalar degrees of freedom.

The ratio between the gauge invariant scalar metric potentials $\phi$ and $\psi$ is used as an indicator of the anisotropic stress of $X^\mn$, see section \ref{sec:megastress}. In addition, the contribution from $X^\mn$ to the Poisson equation is absorbed into a time and scale dependent modification of Newton's constant, defined as $G_{\rm eff}\equiv G(1+\Delta\rho_X / \Delta\rho_m)$, where $\Delta\rho_X$ and $\Delta\rho_m$ are the density perturbations in the comoving orthogonal gauges of matter and dark energy, respectively. The nonlinear counterpart to this approach in the framework that we propose is straightforward: the anisotropic stress is measured through the second equation of \eqref{eq:fullconsistency} and the modified Newton's constant is given by $G_{\rm eff} = G(1+\der_X / \der_m)$, where now $\der_X$ and $\der_m$ are the fluctuations measured in the normal frame and we can write
\be
G_{\rm eff} / G = \der_G / \der_m \,.
\ee
$G_{\rm eff}$ is usually defined in Fourier space. Notice however that when taking $G_{\rm eff}$ to be a ratio between perturbed quantities it makes a difference if it is defined in real or Fourier space. On the other hand, our consistency relations are defined in real space and can directly be translated into Fourier space.

\subsection{Metric perturbations and the normal frame}  \label{sec:metric}

In this section we provide the connection between the covariant approach and second order metric perturbations. We choose to work in the generalised longitudinal gauge, where the spatial components of the normal frame vanish and there is zero shear. For a description of the second order longitudinal gauge for scalar quantities see for instance \cite{Malik:2008im}\,\footnote{In ref.\ \cite{Malik:2008im} the normal vector and the kinematic quantities are given without specifying the gauge. Note that in our notation the roles of $\phi$ and $\psi$ are swapped with respect to theirs.}. 

The homogeneous and isotropic background evolution is characterised by
\be
\br_G = 3 M_P^2 a^{-2} \H^2 &=& \br_m + \br_X
\label{eq:fried_constr}\\
\bp_G = - M_P^2 a^{-2} \Big( 2\H' + \H^2 \Big) &=& \bp_m + \bp_X
\label{eq:fried_dyn}
\ee
where the prime is the conformal time derivative and the conformal Hubble parameter is $\H\equiv a'/a$.  We can define the \emph{geometrical or total equation of state parameter} as
\be
w_G \equiv \frac{\bp_G}{\br_G} = -\frac{1}{3}\Big(1 + 2\frac{\H'}{\H^{2}} \Big) \, .
\ee
Measurements of the background evolution via standard candles and rulers only give $\H(a)$ and not $\br_m$. As a consequence, $\br_G=\br_m+\br_X$ and $\bp_G=\bp_m+\bp_X$ cannot be disentangled without further assumptions \cite{Kunz:2007rk}.

Let us focus on scalar perturbations only, even at second order\footnote{Clearly, this treatment is not complete as second order vectors and tensors are sourced by products of first order scalars and tensors. Since the scalar sector dominates, we assume, for the scope of this work, that the effect of vectors and tensors is negligible.}. The shear seen by the normal frame vanishes when the off-diagonal parts of the metric are set to zero by choosing the gauge functions appropriately. At first and second order we are only left with the usual scalar potentials in the diagonal of the metric, $\psi\equiv\psi_1+\frac{1}{2}\psi_2$ and $\phi\equiv\phi_1+\frac{1}{2}\phi_2$, where the subscripts indicate the order of the perturbation. These metric perturbations coincide with the Bardeen potentials in the longitudinal gauge. For practical reasons that will be clear later on, we choose to work with the variables
\be
\vf_n \equiv \frac{1}{2}\(\phi_n+\psi_n\) \,, \qquad
\Pi_n \equiv \phi_n-\psi_n
\ee
where again the subscript $n$ denotes the order of the perturbation. In the following, wherever this label is omitted we mean the first plus the second order, according to the following convention: $\vf\equiv \vf_1+\frac{1}{2}\vf_2$. The perturbed metric then reads
\be
g_{\mu\nu}dx^\mu dx^\nu = a^2\left\{ -\(1+2\vf-\Pi\)\,d\eta^2
  + \(1-2\vf-\Pi\)\,\delta_{ij} dx^i dx^j \right\}\,.
\ee
The so called Weyl potential, $\vf$, is related to the WL observables and the ISW effect in the Cosmic Microwave Background (CMB). The potential $\Pi$ is related to the anisotropic stress of the theory and quantifies the mismatch between the original potentials $\psi$ and $\phi$. This potential, that is often called \emph{gravitational slip}, is not directly related on its own to specific observables, although it can in principle be reconstructed from combining WL observables with galaxy clustering, as we discuss briefly in section \ref{sec:biasonbias}, or with peculiar velocity measurements, e.g.\ from redshift-space distortions; see also \cite{Amendola:2007rr, Song:2008vm, Song:2008xd}.

In these metric variables, the normal vector on constant conformal time hypersurfaces reads
\be
n^\mu &=& a^{-1}\,\delta^\mu_0\, \left\{1 -\vf +\frac{1}{2}\Pi
  +\frac{3}{2}\(\vf_1-\frac{1}{2}\Pi_1\)^2 \right\}\\
  n_\mu &=& -a\,\delta^0_\mu\, \left\{ 1 +\vf -\frac{1}{2}\Pi 
  -\frac{1}{2}\(\vf_1-\frac{1}{2}\Pi_1\)^2 \right\}
\ee
and the acceleration and isotropic expansion experienced by $n^\mu$ can be found in appendix \ref{app:kinematic}. The shear and the vorticity vanish in this gauge by construction. The relative velocity between the normal frame and the matter rest frame is defined in \eqref{eq:relveldef} and thus satisfies the equation $v_m^\mu = \(1-v_m^2\)^{1/2}u_Y^\mu-n^\mu$. This can be solved perturbatively to find
\be
v_m^\mu = a^{-1}\,\delta_i^\mu\,  V_m^i \,,\qquad
v_{m\,\mu} = a\, \delta_\mu^i \[ V_{m\,i} - (2\vf_1+\Pi_1) V_{m1\,i} \]
\label{eq:relvelpert}
\ee
where $V_{m\,i}=V_m^i=V_{m1}^i+\frac{1}{2}V_{m2}^i$ is the usual velocity perturbation of the matter 4-velocity $u_m^\mu = a^{-1}(V_m^0,\ V_m^i)$, which satisfies $u_m^2=-1$. Finally, the perturbed fluid variables are written in the form $\rho=\br+\der_1+\frac{1}{2}\der_2$ (and analogously for the pressure, energy flux and anisotropic stress perturbations).

Notice that the definition of the perturbation variables depends on the observer frame. Therefore, at second order our perturbation variables are non trivially related to the variables in the usual formalism where the rest frame fluid variables are chosen. As usual, the density and pressure perturbations are not gauge invariant. One may define gauge invariant variables by taking the covariant spatial gradients of the density and pressure \cite{Bruni:1992dg}. The reason is that variables that vanish at the background level are gauge independent at the lowest non-vanishing perturbative order -- a consequence of the Stewart-Walker lemma \cite{Stewart:1974uz}. This also means that $\dep_{m2}$ is gauge independent because $\bar p_m=0=\dep_{m1}$, and analogously for the anisotropic stress. In fact, $\dep_{m2}$ and $\depi_{m2}$ coincide with particular gauge invariant combinations in the longitudinal gauge, see e.g.\ \cite{Hwang:2005hd}.

\subsection{Consistency relations at first order} \label{sec:linear}

We now compute the geometrical fluid variables at linear level by decomposing $G_\mn$ in the normal frame. Focusing on the late time evolution of $\Lambda$CDM, the absence of pressure and anisotropic stress perturbations in matter and the form of $X^\mn=-M_P^2\Lambda g^\mn$, lead to the simple linearised versions of the consistency relations \eqref{eq:fullconsistency}. These linear relations reflect the fact that the two scalar potentials are equal and their evolution is given by the well-known Bardeen equation for $\vf_1$, which we can solve analytically.

The geometrical fluid variables in the normal frame for linear perturbations are found to be
\be
\der_{G1} &=& -2\bar{\rho}_G \left( \vf_1 + \frac{\vf_1'}{\H} - \frac{\nabla^2 \vf_1}{3\H^2}
  -\frac{1}{2} \Pi_1 + \frac{\Pi_1'}{2 \H} - \frac{\nabla^2 \Pi_1}{6 \H^2} \right)
  \label{eq:derG1}
\\
\dep_{G1} &=& \frac{2}{3} \bar{\rho}_G \left(  -3 w_G \vf_1 + 3 \frac{\vf_1'}{\H} + \frac{\vf_1''}{\H^2}
  +\frac{3}{2}w_G \Pi_1 + \frac{\Pi_1'}{2\H} + \frac{\Pi_1''}{2\H^2} - \frac{\nabla^2\Pi_1}{3\H^2} \right)
  \label{eq:depG1}
\\
\deq_{G1}^\mu &=& - \frac{2}{3} \bar{\rho}_G\,\frac{\delta^{\mu i}}{a\H}\, \partial_i \[ \vf_1 +\frac{\vf_1'}{\H}-\frac{1}{2}\Pi_1 +\frac{\Pi_1'}{2\H} \]
  \label{eq:deqG1}
\\
\depi_{G1}^\mn &=& \frac{1}{3} \bar{\rho}_G  \,\frac{\delta^{i\mu}\delta^{j\nu}}{a^2\H^2} \,  \Big(\partial_i\partial_j -\frac{1}{3}\delta_{ij}\nabla^2\Big)
\Pi_1 \, .
  \label{eq:depiG1}
\ee
Well after matter-radiation equality, the matter can be modelled as pressureless dust and the perturbations in the radiation can be neglected. At first order, we have from (\ref{eq:frametransform_pm}) and (\ref{eq:frametransform_pim}) that $\dep_{m1}=0$ and $\depi_{m1}=0$, while $\der_{m1}=\br_m\delta_{m1}$ and $\deq_{m1}^\mu=\br_m V_{m1}^i a^{-1} \delta^\mu_i$, where $\delta_{m1}$ and $V_{m1}^i$ are the first order density contrast and velocity perturbation, respectively. Notice two interesting consequences: first, if the dark energy is not a cosmological constant, its isotropic and anisotropic pressure perturbations, $\dep_{X1}$ and $\depi_{X1}$, can directly be inferred from the metric potentials via (\ref{eq:depG1}) and (\ref{eq:depiG1}), respectively. Conversely, we can also conclude that for any metric inferred from observations there exists a pressure perturbation $\dep_{X1}$ and an anisotropic stress $\depi_{X1}$, given by the equations above, that describe a dark energy component which is able to reproduce the observations. Second, if $\Lambda$CDM is the correct model, the geometric pressure and anisotropic stress perturbations vanish identically and the exact consistency relations \eqref{eq:fullconsistencyLCDM} reduce to
\be
\Lambda{\rm CDM}\quad\Rightarrow\qquad \Pi_1 = 0 \qquad {\rm and} \qquad
\vf_1'' + 3\H \vf_1' -3 w_G\H^2 \vf_1 = 0 \,.
\label{eq:lin_cond}
\ee
The resulting equation for $\vf_1$ is the well-known Bardeen equation for dust after matter-radiation equality. Note that no spatial derivatives are present, which means that $\vf_1$ evolves in a scale-independent way. After matter-radiation equality, we can write the geometric equation of state in $\Lambda$CDM as $w_G=-\Omega_\Lambda/(\Omega_\Lambda+\Omega_m a^{-3})$, where $\Omega_i$ is the present energy fraction in the $i$-component. Then, the equation for $\vf_1$ can be solved analytically, see appendix \ref{app:pertevol}. Neglecting the decaying mode, which soon becomes irrelevant, we find in Fourier space
\be
\vf_1 = C_\Vk {\cal F}\!\(-\alpha^3\) 
\label{eq:lin_phi}
\ee
where $\alpha\equiv (\Omega_\Lambda/\Omega_m)^{1/3}a$ is the scale factor normalised to 1 at $\Lambda$-matter equality, and ${\cal F}(y)\equiv {}_2F_1(1/3,\,1;\, 11/6;\, y)$ is the ordinary hypergeometric function. ${\cal F}(-\alpha^3)$ evolves from $1$ in matter domination to about $0.76$ today and describes the suppression of the growth of perturbations due to the cosmological constant. The integration constant $C_\Vk$ is fixed by the initial conditions that are set well after radiation-matter equality. It can be estimated analytically by solving the perturbation equations of a radiation-matter mixture from the end of inflation until matter domination, neglecting the anisotropic stress of radiation, see e.g.\ equations (2.216) and (3.65) of ref.~\cite{Durrer:2008aa}\footnote{The coefficient for large $k$ is actually given as $20$ in ref.~\cite{Durrer:2008aa}, but we find numerically that the correct coefficient is $3$.}
\be \label{ck}
C_\Vk = -\frac{3}{5}\sqrt{ 2\pi^2 k^{-3} {\cal P}_\zeta(k) }\times \left\{
\begin{array}{ll}
1 & \qquad {\rm for}\ \ k \ll k_s
\vspace{1mm}
\\
3(k_s/k)^2\,\ln (k/k_s) & \qquad {\rm for}\ \ k \gg k_s
\end{array} \right. 
\ee
where ${\cal P}_\zeta(k) = A_s (k/k_p)^{n_s-1}$ is the  primordial dimensionless power spectrum of the comoving curvature perturbation $\zeta$, with $A_s\approx 2\times 10^{-9}$ and $n_s\approx 0.96$ at $k_p=0.05/\Mpc$.  We have introduced the sound horizon at matter-radiation equality, $k_s\equiv \sqrt{3}/\eta_{\rm eq}$, with conformal time at equality, $\eta_{\rm eq}=2(2-\sqrt{2})/\H_{\rm eq}$, and $\H_{\rm eq}=H_0\Omega_m\sqrt{2/\Omega_r} \approx 0.010\,h/\Mpc$, such that $k_s\approx 0.015\,h/\Mpc$. 

We can derive the matter density contrast and the velocity perturbation from $\vf_1$ by means of (\ref{eq:derG1}) and (\ref{eq:deqG1}), see appendix \ref{app:pertevol} for details. In the next section we will use the velocity perturbation to compute the effective matter pressure and anisotropic stress. We find
\be
V_{m1,\Vk}^i = -i {\cal V}(a)\frac{k_i}{\H_0} C_\Vk
\label{eq:Vm1}
\ee
which is valid on all scales. Its time evolution is given by
\be
{\cal V}(a) \equiv \frac{2}{3} 
  \[\frac{a(1+\alpha^3)}{\Omega_m}\]^{1/2} \left[ {\cal F}\!\(-\alpha^3\)
  -3\alpha^3 \tilde{\cal  F}\!\(-\alpha^{3}\)  \right]
\label{eq:velevol}
\ee
where $\tilde{\cal F}(y)\equiv (d/dy){\cal F}(y)$.

Let us notice that the effortless way in which the Bardeen equation was derived here is due to the application of the consistency relations of the geometric pressure that we introduced in the last section. Next we will see how this is generalised at second order and how the effective matter pressures come into play.

\section{Second order perturbations} \label{sec:nonlinear}

In this section, we derive the consistency relations in second order perturbation theory. When going beyond linear perturbation theory, the choice of the observer frame becomes important. As discussed in section \ref{sec:framechange}, the transformation from the rest frame to the normal frame gives rise to an effective pressure and anisotropic stress proportional to the square of the relative velocity, even for a pressureless perfect fluid. We compute and quantify these contributions in $\Lambda$CDM such that they can be properly taken into account when applying the second order consistency relations.

\subsection{Effective pressures}  \label{sec:effectivepress}

\begin{figure}[tb]
\centering
\includegraphics[width=0.48\textwidth]{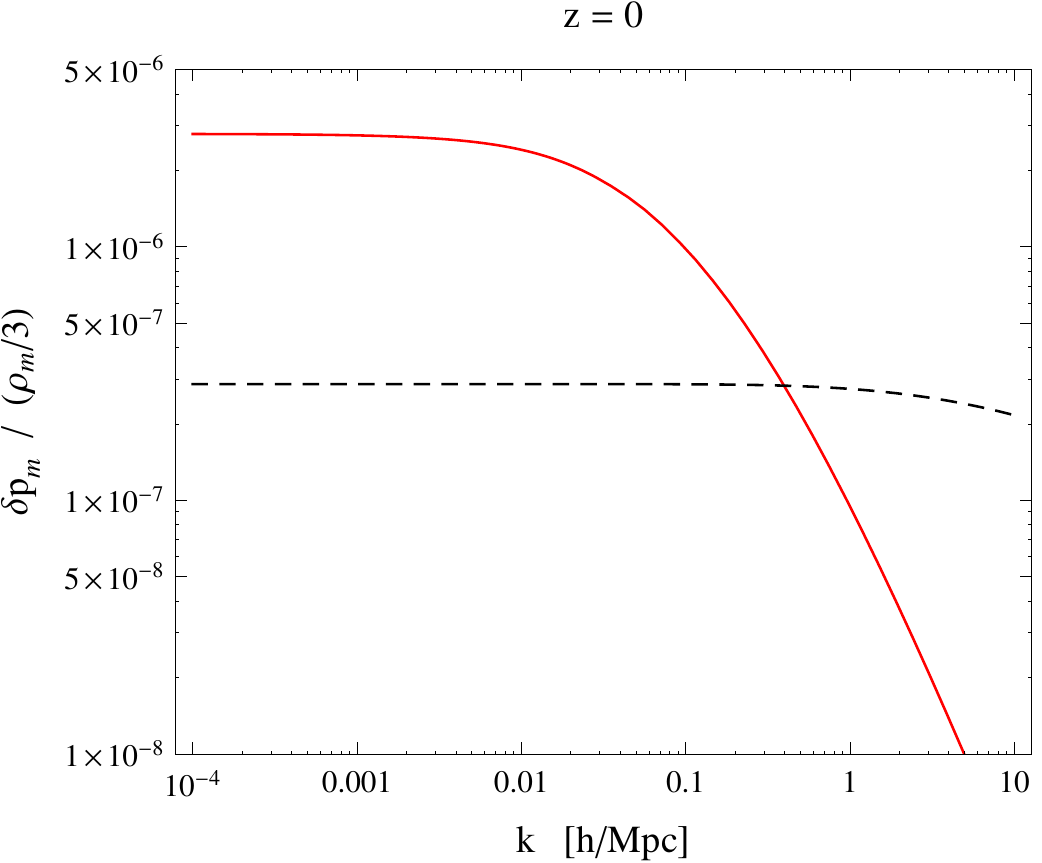}\quad
\includegraphics[width=0.48\textwidth]{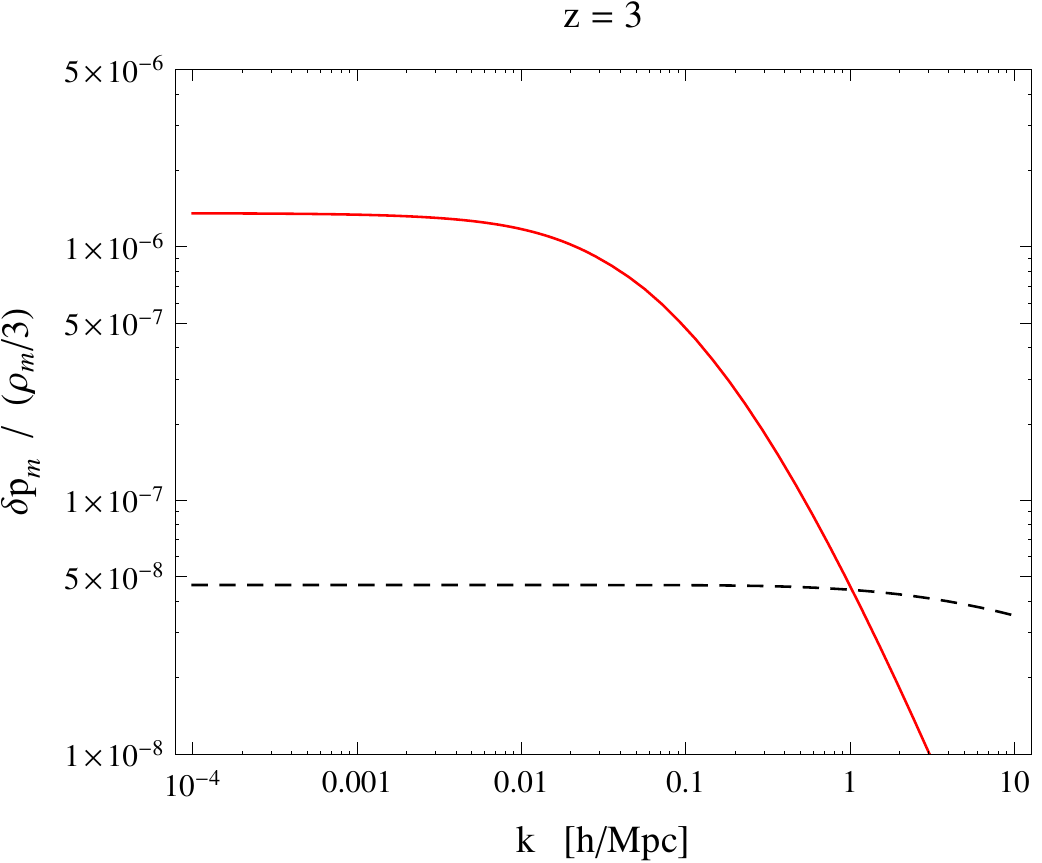}
\caption{\label{fig:pm}
The matter pressure perturbation in the normal frame is shown for a typical $\Lambda$CDM cosmology at redshift $z=0$ (left panel) and $z=3$ (right panel) as a function of $k$. The second order contribution from $V_{m1}^2$ (solid red) falls below the leading third order contribution from $\delta_{m1}V_{m1}^2$ (dashed black) around $k\approx 0.4\,h/\Mpc$ at $z=0$ and around $k\approx 1\,h/\Mpc$ at $z=3$.}
\end{figure}

The effective pressure and anisotropic stress induced by the change of frame are given in exact form in (\ref{eq:frametransform_pm}) and (\ref{eq:frametransform_pim}). We use the perturbed relative velocity from \eqref{eq:relvelpert} to find
\be \label{deltape}
\dep_{m2} &=& \frac{2}{3}\br_m V_{m1}^i V_{m1\,i}
\\  \label{deltapi}
\depi_{m2}^\mn &=& 2\br_m \left\{ V_{m1}^i V_{m1}^j -\frac{1}{3}\delta^{ij} V_{m1}^k V_{m1\,k}
 \right\} \, a^{-2}\delta^\mu_i\delta^\nu_j \,.
\ee
The real-space product in~(\ref{deltape}) becomes a convolution in Fourier space such that
\be
\dep_{m2}(\Vk) = \frac{2\br_m {\cal V}^2}{3 \H_0^2}
  \int\frac{d^3q}{(2\pi)^3}\, \(q^2-\Vk\cdot\Vq\) C_{\Vk-\Vq}\, C_\Vq
\ee
where we used \eqref{eq:Vm1} to express the velocity perturbation in terms of the matter era amplitude of the Weyl potential, $C_\Vk$. For large scales, $k\ll k_s$, it is easy to estimate the convolution using the approximation \eqref{ck} for $C_\Vk$. The angular integration becomes trivial and the radial one can be split at $k_s$. We find
\be \label{eq:depimest}
\dep_{m2}(k\ll k_s) &\simeq& \frac{18}{75} \br_m {\cal V}^2 A_s \bigg(\frac{k_s}{k_p}\bigg)^{n_s-1}
  \bigg(\frac{k_s}{\H_0}\bigg)^2 \( \frac{1}{1+n_s} + \frac{18}{3-n_s} \)
\ee
which provides a good estimate of the true numerical result that we compute using output from a modified version of the CMB-Boltzmann code CAMB \cite{Camb,Lewis:1999bs}. For the small-scale behaviour, $k\gg k_s$, we find analytically that $\dep_{m2}$ drops off like $\ln(k/k_s)(k/k_s)^{n_s-3}$ which also matches the behaviour of the numerical results. However, due to the involved angular integration it is not straightforward to derive a simple expression for the amplitude on small scales and it is beyond the scope of this work to investigate this in more detail. In figure \ref{fig:pm} we show the numerical result for the effective matter pressure perturbation at redshifts $0$ and $3$. To get an idea at which wavenumbers a second order calculation is not sufficient anymore, we also compute the leading third order correction, the term proportional to $\delta_{m1}V_{m1}^2$.\footnote{We plot this third order term with the sole aim of showing, by comparison with the complete second order, at which intermediate (mildly nonlinear) scales, higher order correction terms can start to be relevant. However, our analysis is not meant to be a full perturbative treatment at those scales, which would probably require the use of resummation techniques (e.g.\ \cite{Crocce:2005xy} and \cite{Pietroni:2008jx}).} The second order contribution falls below the leading third order contribution around $k\approx 0.4\,h/\Mpc$ at $z=0$ and around $k\approx 1\,h/\Mpc$ at $z=3$.

\begin{figure}[tb]
\centering
\includegraphics[width=0.48\textwidth]{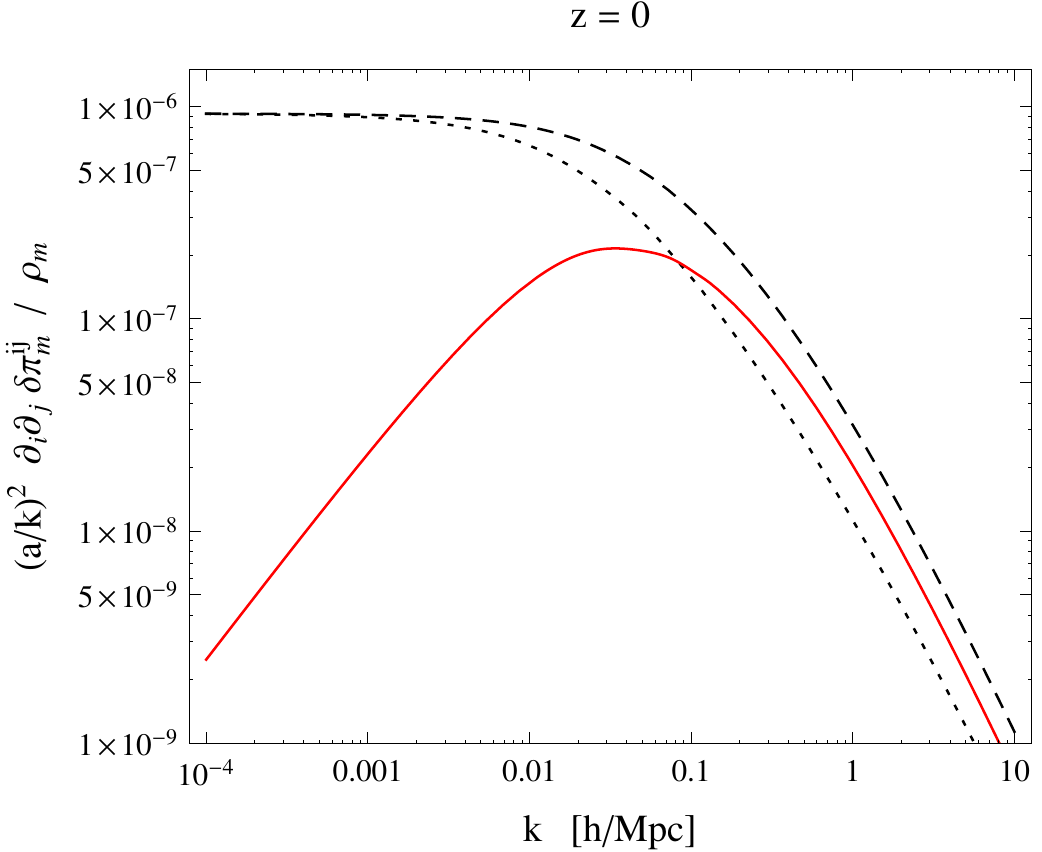}\quad
\includegraphics[width=0.48\textwidth]{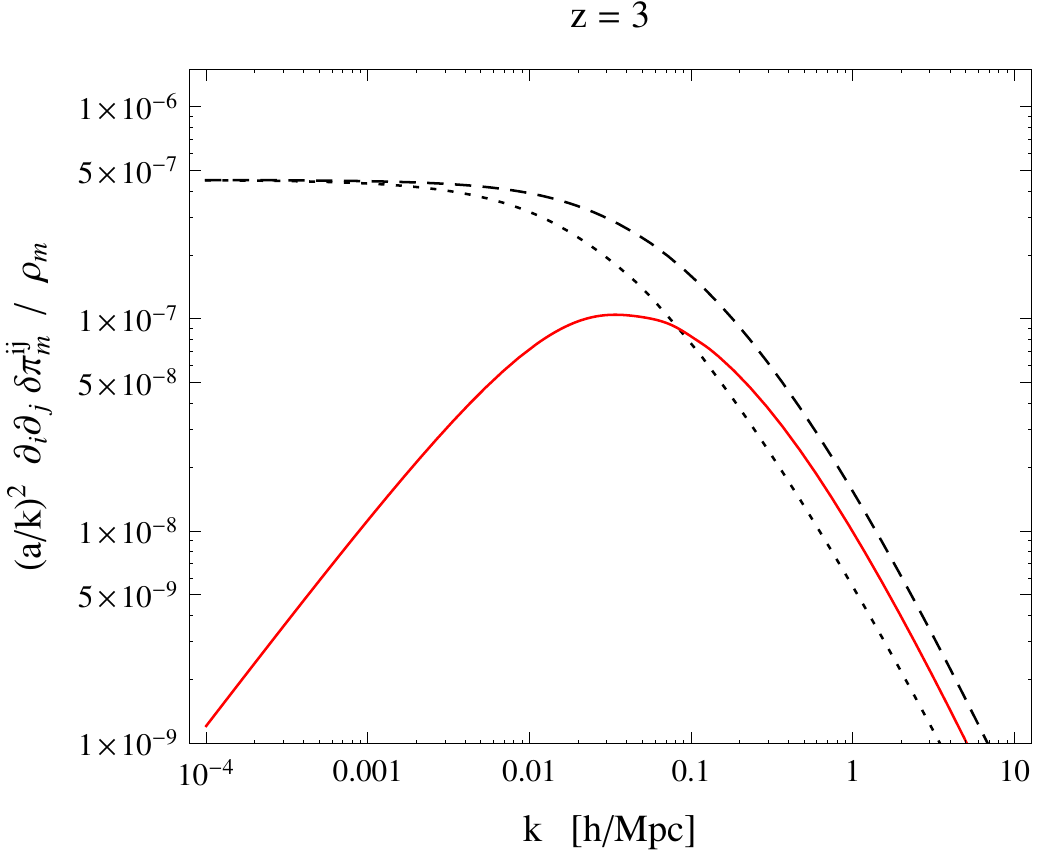}
\caption{\label{fig:pim}
The matter anisotropic stress in the normal frame is shown for a typical $\Lambda$CDM cosmology at redshift $z=0$ (left panel) and $z=3$ (right panel) as a function of $k$. On large scales, $k\ll k_s$, the term $\propto V_{m1}^2$ (dashed black) is partly canceled by the contribution $\propto k^{-2} \partial_i\partial_j( V_{m1}^i V_{m1}^j )$ (dotted black) which is negative, such that their sum (solid red) falls off like $k/k_s$.}
\end{figure}

For the anisotropic stress we take a double spatial divergence to construct a scalar quantity
\be
\frac{a^2}{k^2}\,\partial_i\partial_j\depi_{m2}^{ij}(\Vk) = \frac{2\br_m {\cal V}^2}{3 \H_0^2}
\int\frac{d^3q}{(2\pi)^3}\,\[q^2+2\Vk\cdot\Vq-3(\hat{\Vk}\cdot\Vq)^2\]  C_{\Vk-\Vq} \, C_\Vq
\ee
where $\hat{\Vk}\equiv \Vk/k$. We compute the matter anisotropic stress numerically and show the resulting behaviour in $k$ at redshifts $0$ and $3$ in figure \ref{fig:pim}. When computing this convolution we realise that on large scales, $k\ll k_s$, the contributions from the term $\propto \partial_i \partial_j (V_{m1}^i V_{m1}^j)$  in \eqref{deltapi} is negative and partly cancels the term $\propto\nabla^2 V_{m1}^2$. We analysed the small-$k$ behaviour analytically using a number of approximation schemes and found $k^{-2} \partial_i\partial_j\depi_{m2}^{ij}(k\ll k_s) \propto k/k_s$, which correctly reproduces the numerical results. Again, an exact analytical estimate of the small-$k$ limit is difficult to achieve and not helpful for the discussion presented here. On small scales, $k\gg k_s$, the behaviour of the anisotropic stress is dominated by the term $\propto\nabla^2 V_{m1}^2$ and therefore the scaling is also $\propto \ln(k/k_s)(k/k_s)^{n_s-3}$ as discussed above.

\begin{figure}[tb]
\centering
\includegraphics[width=0.48\textwidth]{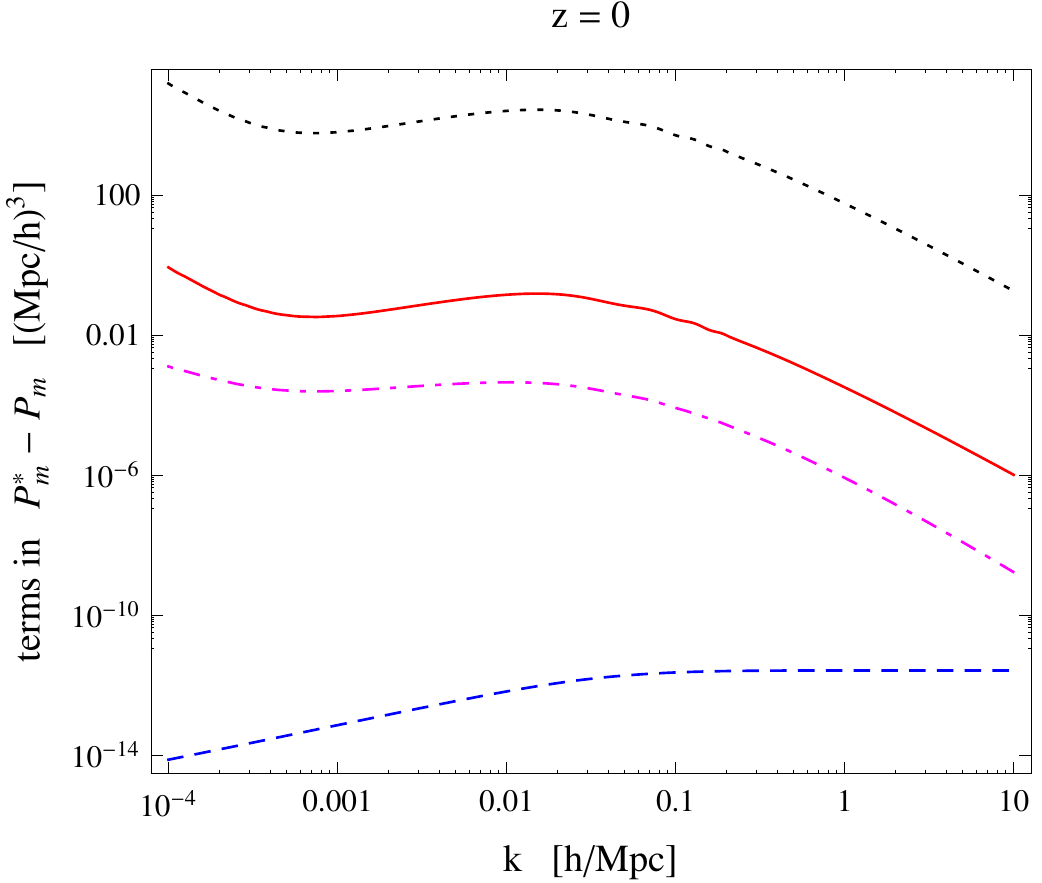}\quad
\includegraphics[width=0.48\textwidth]{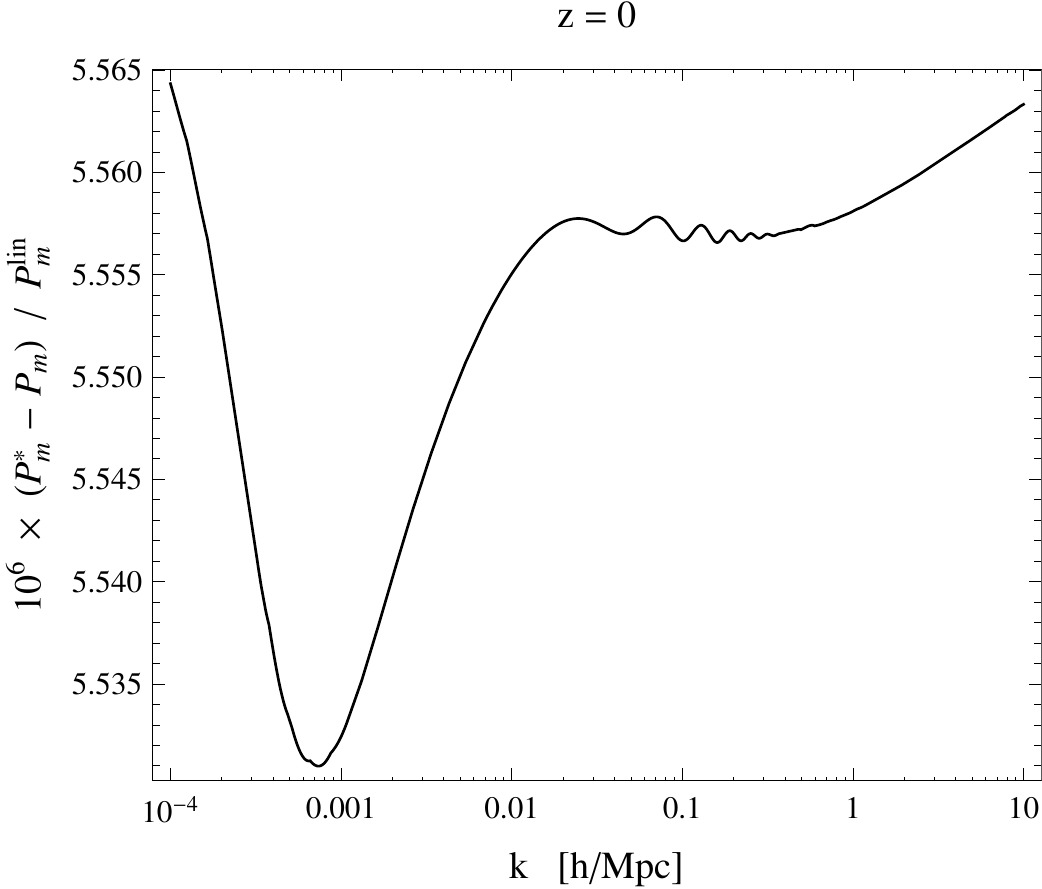}
\caption{\label{fig:pmdiff}
The difference between the comoving and the normal frame matter power spectrum at second order in the longitudinal gauge, $P_m^*-P_m$, is shown for a typical $\Lambda$CDM cosmology at redshift $z=0$ as a function of $k$. In the left panel the different contributions are plotted: $(2\nu-\nu^2)P_m^{\rm lin}$ (solid red), $2P_{\rm cross}$ (dot-dashed magenta), $\nu^2-P_{\rm auto}$ (dashed blue). For comparison we show the linear matter power spectrum $P_m^{\rm lin}$ in dotted black. The right panel shows the relative difference, $(P_m^*-P_m)/P_m^{\rm lin}$, scaled by a factor of $10^6$.}
\end{figure}

Finally, let us study the difference between the density perturbations in the normal frame and the comoving frame. As we saw in section \ref{sec:framechange}, the difference between the two point correlation functions of the density fields is due to the relative difference of the mean densities (\ref{eq:nu}), and the pressure-pressure and pressure-density correlations, see \eqref{eq:fullcorrelfunc}. We compute this difference up to second order and express it in terms of the power spectra to find
\be \label{eq:pmdiff}
P_m^* - P_m = \(2\nu -\nu^2\) P_m^{\rm lin} - 2P_{\rm cross} -\left[\nu^2-P_{\rm auto}\right]
\ee
where 
\be
(2\pi)^3 \delta_D(\Vk-\Vk') P_m^{\rm lin}(k) &\equiv& \la \delta_{m1}(\Vk)\, \delta_{m1}^*(\Vk') \ra 
\\
(2\pi)^3 \delta_D(\Vk-\Vk') P_{\rm cross}(k) &\equiv& \la \delta_{m1}(\Vk)\, \,[V_{m1}^iV_{m1\,i}]^*(\Vk') \ra 
\\
(2\pi)^3 \delta_D(\Vk-\Vk') P_{\rm auto}(k) &\equiv& \la [V_{m1}^iV_{m1\,i}](\Vk)\, \,[V_{m1}^iV_{m1\,i}]^*(\Vk') \ra  \, .
\ee
The linear densities coincide in the normal and comoving frames, and we find that $\nu = [V_{m1}^iV_{m1\,i}](k=0)$ is the velocity dispersion. This gives a clear physical interpretation for the difference between the two frames: the geometric matter pressure seen by the normal frame is due to its velocity dispersion. In figure \ref{fig:pmdiff} we show the different terms in \eqref{eq:pmdiff} and the relative difference between the matter power spectra in the two frames in a typical $\Lambda$CDM model. It is evident that the term $2\nu P_m^{\rm lin}$ is the dominant contribution. In the present epoch we find the velocity dispersion to be $\nu(z=0) \approx 3\times 10^{-6}$ (for $\Lambda$CDM, in units $c=1$, see figure \ref{fig:pim}), a value that corresponds to velocities of $v\approx 500\,$km/s. The fact that $\nu$ is non-zero means that there is a fundamental difference between the comoving matter frame and non-comoving frames. Indeed, the mean densities are related by $\br_m^* = (1-\nu)\br_m$ which means that the background evolution is slightly different in the two frames; different only at the level of second order perturbations. This means that in principle the background needs to be renormalised in order to take this into account, a procedure that belongs into the realm of backreaction calculations and is not of interest for the present work; see e.g.\ \cite{Baumann:2010tm}.

We conclude that the amplitude of $\dep_m$ today is about $10^{-6}\br_m$ on large to intermediate scales and is therefore small but a priori not negligible.  We will compare it with the pressure perturbation in typical dynamical DE models in section \ref{sec:depress}. It is interesting to note that $\dep_m$ does not vanish for $k\to 0$ which implies that, in principle, the large-scale average matter pressure is non-zero and needs to be renormalised. This was also recently pointed out in \cite{Baumann:2010tm}.  The matter anisotropic stress, on the other hand, is found to be vanishing for $k\to 0$. It peaks roughly at $k\sim k_s$ with an amplitude of $\sim 10^{-7}\br_m$. In the next section, we will compare it with the other contributions to the second order gravitational slip, $\Pi_2=\phi_2-\psi_2$, which form the second order consistency relations.

\subsection{Consistency relations at second order} \label{sec:secconsistency}

We calculate the geometrical fluid variables, $\der_{G2}$, $\dep_{G2}$, $\deq_{G2}^\mu$ and $\depi_{G2}^\mn$, as measured in the normal frame up to second order in the metric perturbations. The full expressions for all these quantities are presented in appendix \ref{app:secondordergeometrical}. Here we focus on the geometrical pressure and anisotropic stress, given in \eqref{eq:depG2} and \eqref{eq:depiG2}, that are used to write the second order consistency relations \eqref{eq:fullconsistency}. Here we assume $\Pi_1=0$, but it is straightforward to include a non-zero first order gravitational slip. The anisotropic stress consistency reads
\be
\nabla^4\Pi_2 &=&  \frac{9\H^2}{2\br_G}a^2\partial_i\partial_j\depi_{G2}^{ij}
  -8\vf_1\nabla^4\vf_1 -2(\nabla^2\vf_1)^2 -12\nabla^2(\nabla\vf_1)^2
  +10(\partial_i\partial_j\vf_1)^2 \,.
\label{eq:k4pi}
\ee
The second order gravitational slip, $\Pi_2$, comes in as a source term in the second order Bardeen equation, which follows from the pressure consistency
\be
\frac{\vf_2''}{\H^2} + 3\frac{\vf_2'}{\H} -3w_G \vf_2
&=& 
  \frac{3\dep_{G2}}{2\br_G}
  +\( \frac{\nabla^2\Pi_2}{3\H^2} -\frac{\Pi_2''}{2\H^2} -\frac{\Pi_2'}{2\H} -\frac{3}{2}w_G \Pi_2\)
\nn \\ && 
  + 8\vf_1 \frac{\nabla^2 \vf_1}{3\H^2}
  + \frac{7(\nabla \vf_1)^2}{3\H^2}
  -12w_G \vf_1^2 
  + 8\vf_1 \frac{\vf_1'}{\H} + \frac{(\vf_1')^2}{\HH^2}
   \,.
\label{eq:bardeen2}
\ee
The geometric pressure perturbation and anisotropic stress are given by the effective matter stresses that we computed in the last section and any stresses coming from $X^\mn$:
\be
\dep_{G2} = \dep_{m2} + \dep_{X2}
\,,\qquad
\depi_{G2}^{ij} = \depi_{m2}^{ij} + \depi_{X2}^{ij} \,.
\ee
The consistency of $\Lambda$CDM requires $\dep_{X2}=0=\depi_{X2}^{ij}$. A few comments are in order: first, notice that the second order gravitational slip $\Pi_2$ is purely given by gravitational nonlinearities, $\partial^4\vf^2$--type terms, and the anisotropic stress measured by the normal frame, $\partial_i\partial_j\depi_{G2}^{ij}$. This means that a gravitational slip is generated at second order regardless of the model, even in $\Lambda$CDM where there is no anisotropic stress in the comoving frame. Second, an analogous observation can be made about the source terms of the second order Bardeen equation \eq{eq:bardeen2}. Again, apart from the gravitational nonlinearity terms, the source is given by the pressure and anisotropic stress measured in the normal frame. Finally, it is important to stress that a gravitational slip is generated at second order independently of our choice to work with the normal frame. We chose to use the normal frame because it provides a very direct way of arriving at the above equations and enables us to interpret the contribution of the stress-energy content to the nonlinear source terms.

\begin{figure}[tb]
\centering
\includegraphics[width=0.48\textwidth]{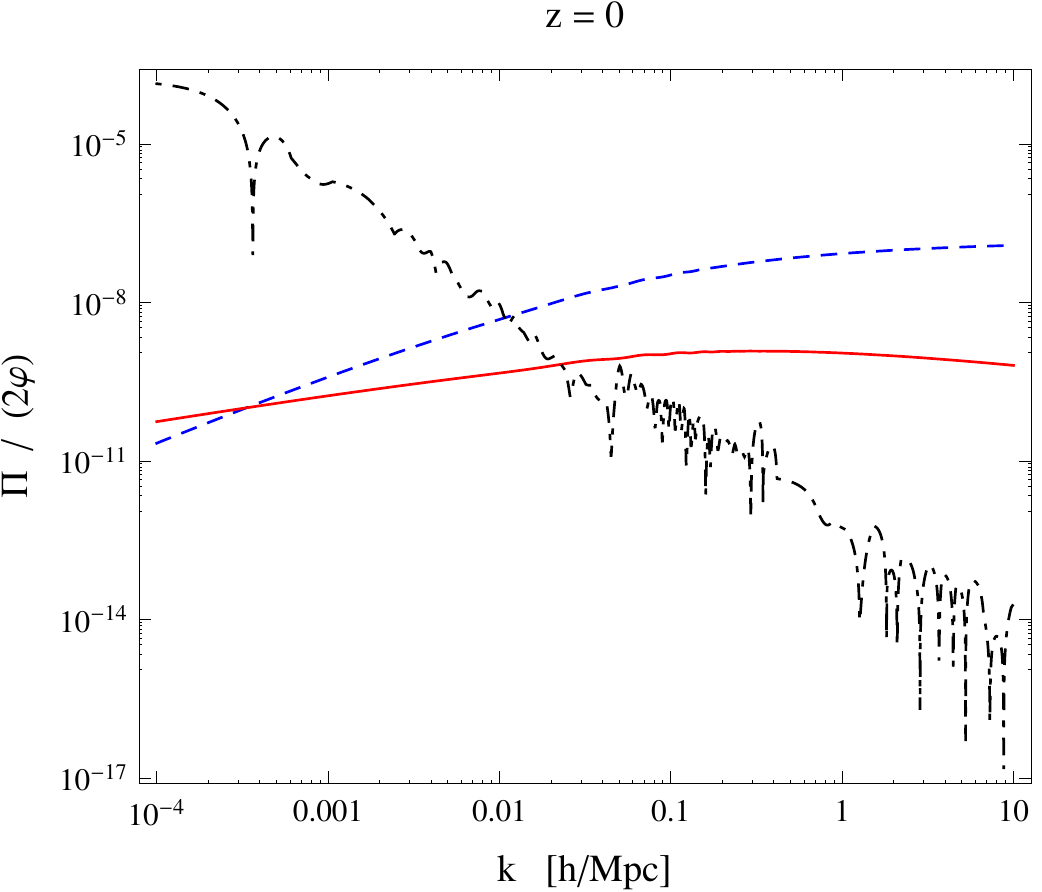}\quad
\includegraphics[width=0.48\textwidth]{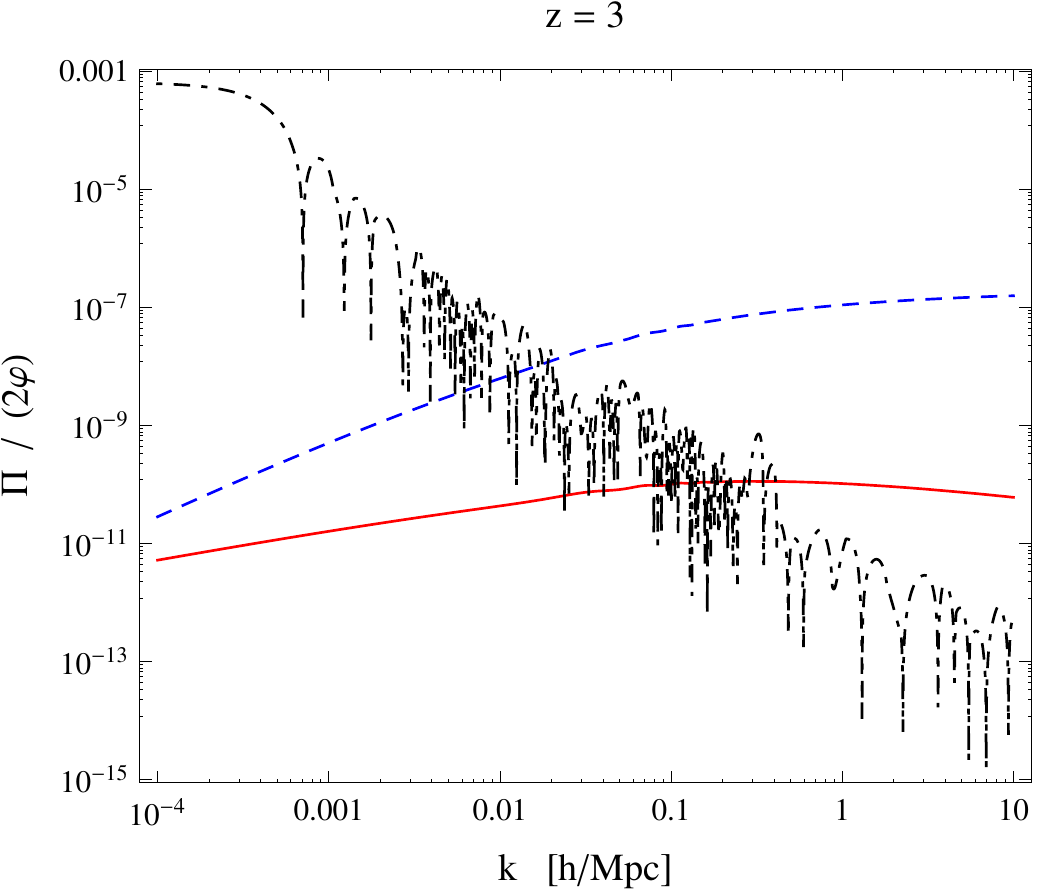}
\caption{\label{fig:piby2phi}
Different contributions to $\Pi/(2\vf)$ are shown for a typical $\Lambda$CDM cosmology at redshift $z=0$ (left panel) and $z=3$ (right panel) as a function of $k$. The $\partial^4\vf^2$--type terms (dashed blue) mostly dominate over the term $\propto \partial_i\partial_j(V_{m1}^iV_{m1}^j)$ (solid red), while on large scales, $k\lesssim 0.01\, h/\Mpc$ the anisotropic stress from photons and massless neutrinos dominates (dot-dashed black).}
\end{figure}

In the case of $\Lambda$CDM, we compute $\Pi_2(k)$ from \eqref{eq:k4pi} by numerically convolving the potential computed with CAMB. In figure \ref{fig:piby2phi} we show the $k$-dependence of the different contributions to $\Pi/(2\vf)$ for a typical $\Lambda$CDM cosmology at redshifts $0$ and $3$ where we compare it to the first order gravitational slip from photon and massless neutrino anisotropic stress (that is not very accurately computed in CAMB at late times and on small scales). The contribution from the $\partial^4\vf^2$--type terms  in \eqref{eq:k4pi} turns out to be larger than the contribution from the matter anisotropic stress, but comes with the opposite sign. Combining the terms, we find that we have to compute
\be
\partial^4\varphi^2 \sim 2 \int_0^\infty dq \int_{-1}^1 d\mu \, \frac{q^2}{(2\pi)^2} \Big[ -8 k q^3 \mu + 5 k^2 q^3 (1+\mu^2) - 6 k^3 q \mu \Big]  C_{|\bf k- \bf q|} C_q .
\ee
For small $k\ll k_s$ we find that the angular integration cancels the $k$ and $k^3$ terms while the $q$ integration results in a vanishing $k^2$ term. The leading order term picks up a logarithmic correction, and we find that $\partial^4\varphi^2 \propto k^4 \ln(k/k_s)$, i.e.\ these terms go to zero somewhat faster than the matter anisotropic stress.

\begin{figure}[tb]
\centering
\includegraphics[width=0.48\textwidth]{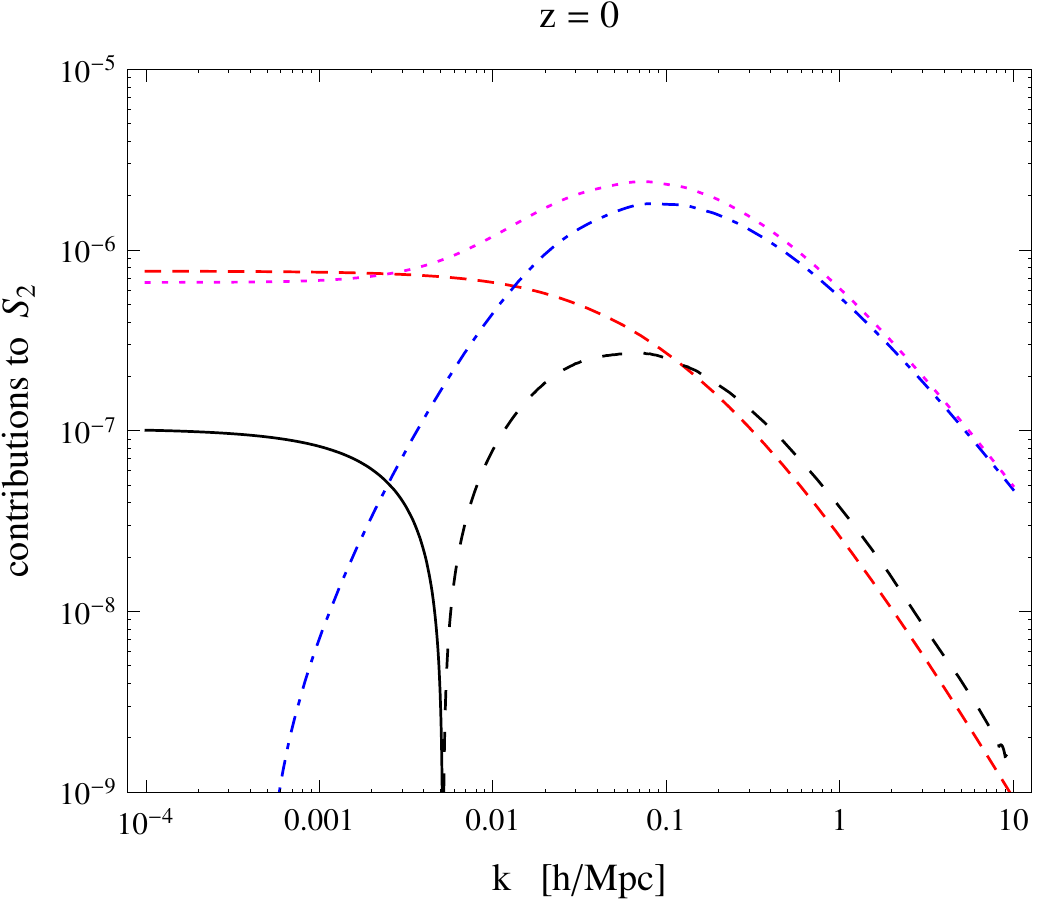}\quad
\includegraphics[width=0.48\textwidth]{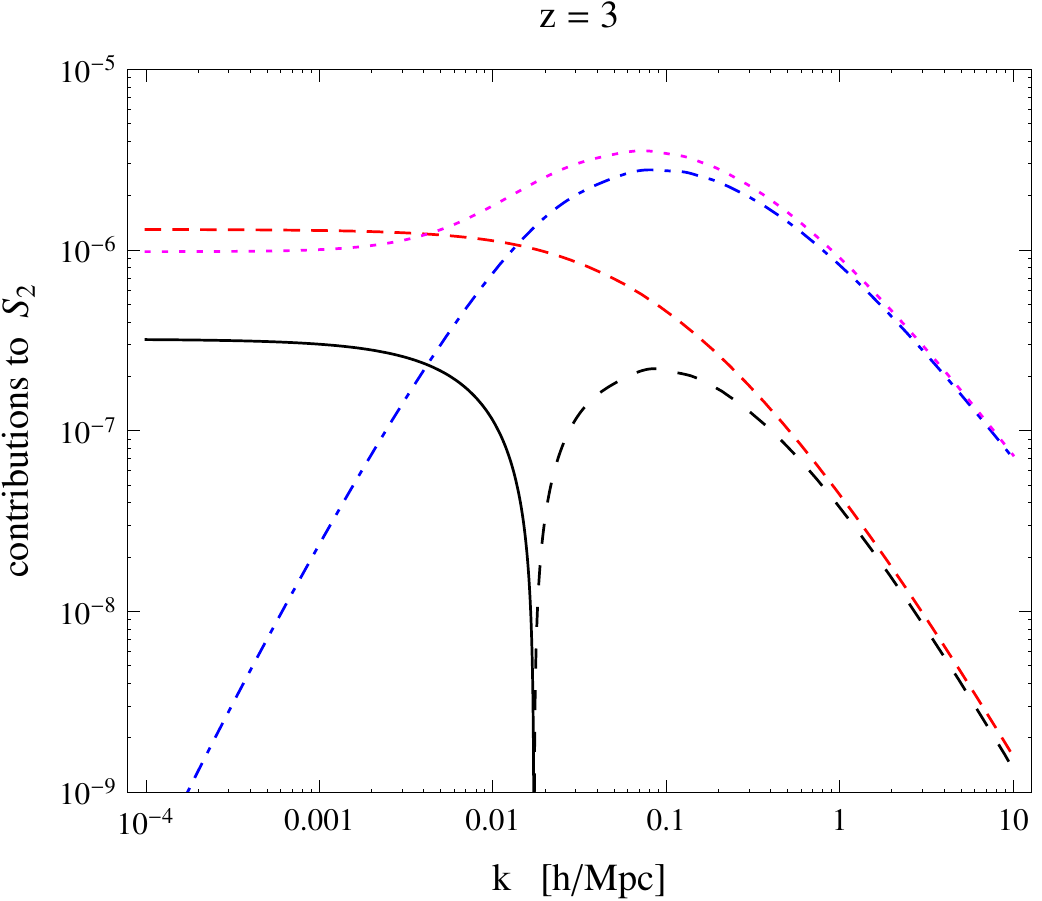}
\caption{\label{fig:sources}
The three different relevant source terms of the second order Bardeen equation are compared for a typical $\Lambda$CDM cosmology at redshift $z=0$ (left panel) and $z=3$ (right panel). The matter pressure source $S_\dep$ (dashed red) and the gravitational slip source $S_\Pi$ (dot-dashed blue) are both positive, while the $\partial^2\vf^2$--type source $S_{\partial^2\vf^2}$ (dotted magenta) is negative. This leads to some cancellations in the total source term (black) which is positive (solid) on large scales, $k\lesssim 0.01\, h/\Mpc$ (depending on redshift), and negative (long dashed) on small scales.}
\end{figure}

Next, let us have a look at the second order Bardeen equation. The source term on the right hand side is a combination of terms from the matter pressure, the gravitational slip that includes the matter anisotropic stress, and further $\partial^2\vf^2$--type terms. At weakly nonlinear scales the terms with time derivatives and those without derivatives can safely be neglected in comparison to the gradient terms. We write the total second order source term as
\be
S_2 &\simeq& S_\dep+S_\Pi+S_{\partial^2\vf^2}
\\
S_\dep &\equiv& \frac{3\dep_{G2}}{2\br_G} = \frac{V_{m1}^2}{1+\alpha^3}
\\
S_\Pi &\equiv& \frac{\nabla^2\Pi_2}{3\H^2} = - \frac{a}{3\Omega_m(1+\alpha^3)}\frac{k^2}{\H_0^2} \Pi_2
\\
S_{\partial^2\vf^2} &\equiv& 8\vf_1 \frac{\nabla^2 \vf_1}{3\H^2}+ \frac{7(\nabla \vf_1)^2}{3\H^2}
\\
&=& -\frac{a{\cal F}(-\alpha^3)^2}{2\Omega_m(1+\alpha^3)} \H_0^{-2}
\int\frac{d^3q}{(2\pi)^3}\, \(q^2+7\,\Vk\cdot\Vq\) C_{\Vk-\Vq}\, C_\Vq \,.
\ee
We compute the convolution in $S_{\partial^2\vf^2}$ numerically using $\vf(k)$ from CAMB for a typical $\Lambda$CDM cosmology at redshifts $0$ and $3$ and compare it to the other two source terms in figure \ref{fig:sources}. We find that  $S_{\partial^2\vf^2}$ is negative while $S_\dep$ and $S_\Pi$ are positive, such that some partial cancellations occur in the total source term, $S_2$. This turns out to be positive and roughly constant on large scales, $k\lesssim k_s$, where it turns negative and starts to decay in the same way as $\dep_{m2}$, so $\propto \ln(k/k_s)(k/k_s)^{n_s-3}$. Using the source term derived here, it is now in principle possible to solve the second order Bardeen equation for $\vf_2$, but we leave this for future work since it is not the aim of this paper. We emphasise however, that the derivation of the Bardeen equation was simple taking the nonlinear consistency relations as a starting point. It is straightforward to generalise this to higher order. Moreover, this makes clear how a possible dark energy pressure, $\dep_{X2}$, would come in as a source term. Due to the subtle cancellation we find, it is likely that the dark energy pressure would dominate the source term on small scales if the dark energy sound speed is not too close to unity and its equation of state parameter is not too close to $-1$. If this is the case, the sign-change and drop-off in $S_2$ moves to smaller scales, see figure \ref{fig:sources}, and $\vf_2$ grows instead of decaying where the source becomes positive and appreciable. Therefore we can expect $\vf_2$ to strongly depend on the dark energy model. Extrapolating from linear theory, we expect that specific combinations of the Weyl potential, $\vf_2$, and the gravitational slip, $\Pi_2$, are directly linked to WL, peculiar velocity and galaxy clustering observables via the Euler and Poisson equations. Therefore, their behaviour on mildly nonlinear scales can provide an interesting smoking gun for non trivial dark energy dynamics that to be tested by combining different cosmological probes. A determination of the dependence of $\varphi_2$ on the dark energy properties and, hence, of its effects on the aforementioned observables, cannot generally be done without resorting to specific models, but the argument above nevertheless shows that these may be used as probes of dark energy.

\section{Applications and discussion}\label{sec:discussion}

\subsection{Dark energy pressure perturbation} \label{sec:depress}

In $\Lambda$CDM the matter pressure $\dep_m$ from \eq{eq:frametransform_pm} is the only source of pressure perturbations well after radiation-matter equality. However, if the late time accelerated expansion is not driven by a cosmological constant but by some form of dark energy in GR, we can expect an intrinsic pressure perturbation $\dep_X$ to appear along with fluctuations in the dark energy component. In this case, according to the first equation in \eq{eq:fullconsistency}, the total pressure perturbation $\dep_G$ is the sum of $\dep_m$ and $\dep_X$. This raises the question of which one of these two terms is more important for generic dark energy models. Whereas the first non-zero contribution to matter perturbations \eq{eq:relvelpert} appears at second order, dark energy perturbations are already present at first order. However, dark energy density perturbations are typically expected to be small in comparison to matter ones, $\delta_X\sim(1+w_X)\delta_m$, and the behaviour of the dark energy pressure perturbation depends on the sound speed. Therefore, it is a priori unclear how matter and dark energy pressure perturbations compare to each other. Here, we compare the pressure perturbation of a perfect fluid dark energy model with the effective matter pressure that we quantified in the last section. We do this to find out whether the matter pressure could, in principle, be mistaken for a dark energy pressure perturbation, or if it has to be taken into account when trying to reconstruct the dark energy properties from its pressure perturbation. As we will see next, the precise answer to this question depends on the wavelength of the fluctuations.

\begin{figure}[tb] 
\centering
\includegraphics[width=0.48\textwidth]{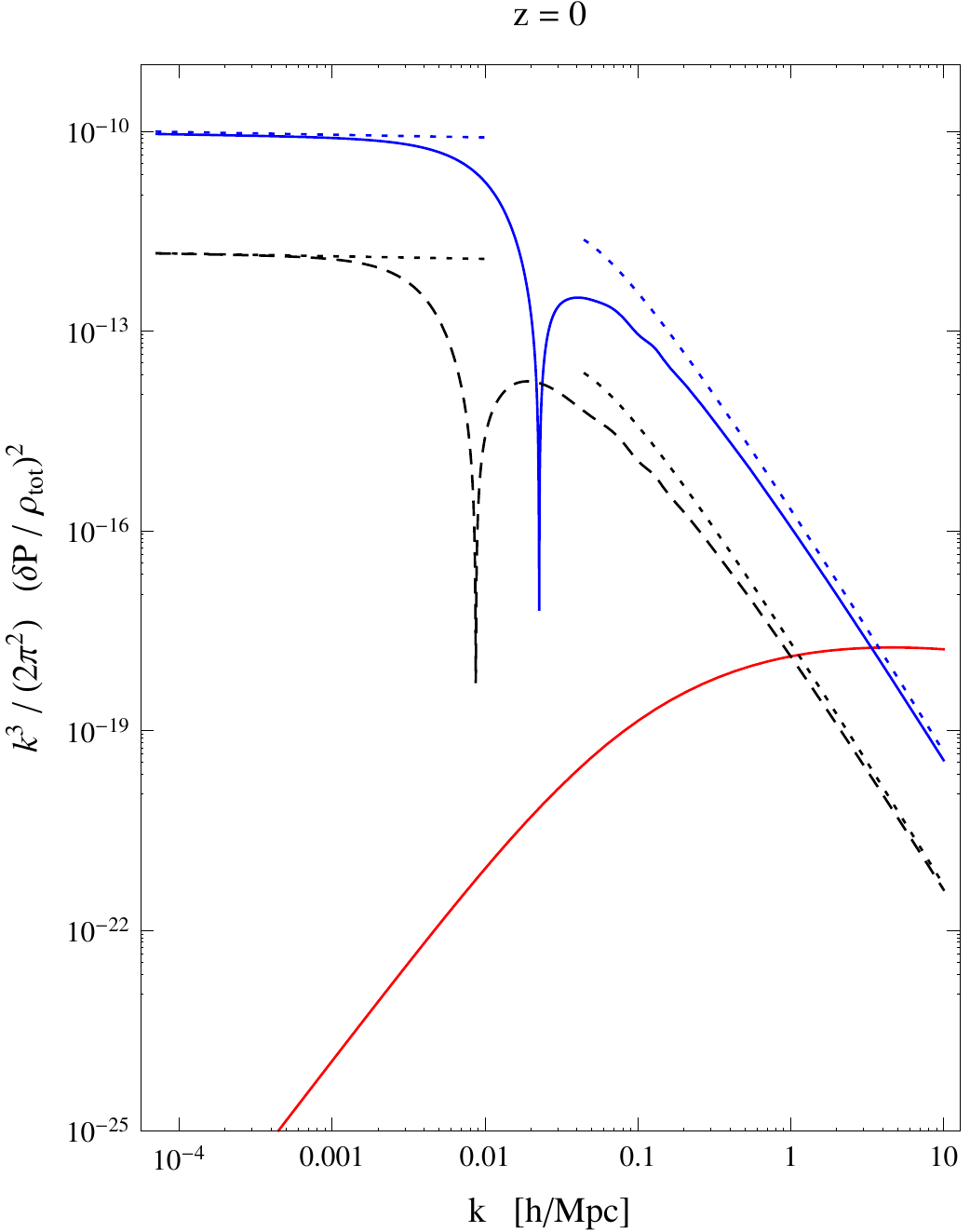}\quad
\includegraphics[width=0.48\textwidth]{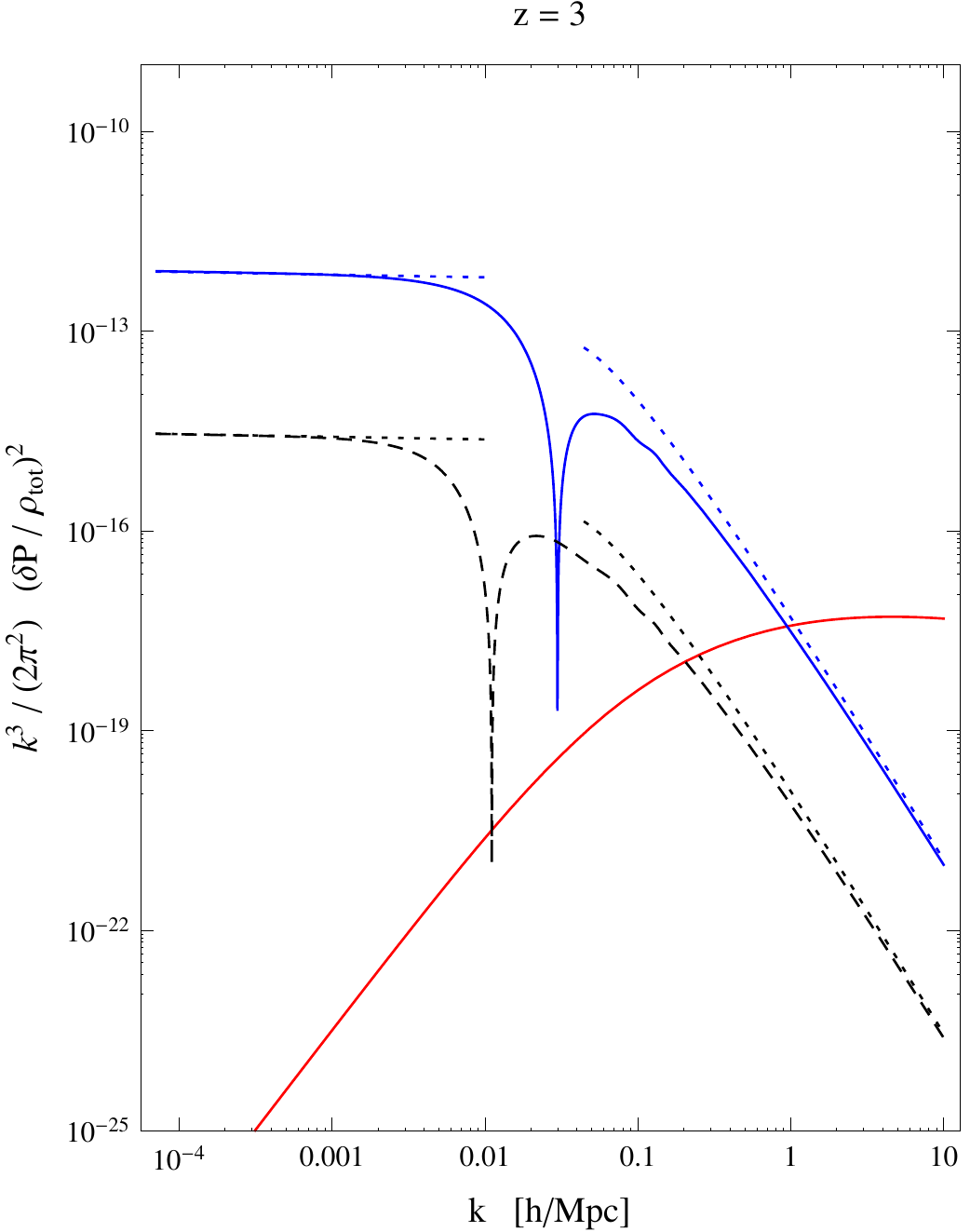}
\caption{\label{presscomp}
Comparison of the dimensionless power spectra of dark energy pressure perturbations and effective matter pressure $\dep_X/(\br_m+\br_X)$ (black \& blue) and $\dep_m/(\br_m+\br_X)$ (red), respectively, at $z=0$ (left panel) and $z=3$ (right panel). For the matter pressure we show the second order result $\propto V_{m1}^2$ (solid red). For the clustering DE we compare two models: $w_X=-0.8$, $\cshat=0.001$ (solid blue) and  $w_X=-0.98$, $\cshat=0.01$ (dashed black). The equation of state, $w_X$, controls the amplitude while the sound speed, $\cshat$, controls the scale at which the spectrum starts falling off on small scales. The dotted asymptotes to the dark energy spectra for small and large scales are given by the analytical approximations \eq{genp2} and \eq{applk}. The intrinsic dark energy pressure is orders of magnitude larger than the effective matter pressure for large scales. However, for small scales, $k\gtrsim 0.3\,h/\Mpc$, the second order matter pressure starts to become relevant depending on the dark energy model parameters and redshift.}
\end{figure}

It is well known that the amplitude of dark energy fluctuations increases as the equation of state parameter $w_X$ deviates from $-1$. In addition, the importance of the dark energy clustering is determined by the sound speed which sets the Jeans length $\sim {c_X}\H^{-1}$ below which pressure support inhibits gravitational collapse. In the limit in which the sound speed equals the speed of light, clustering becomes negligible and therefore virtually impossible to detect. For an incomplete list of references where dark energy perturbations are studied see \cite{Bean:2003fb, Takada:2006xs, Kunz:2006ca, Ballesteros:2008qk, Sapone:2009mb, dePutter:2010vy, Ballesteros:2010ks, Sapone:2010uy, Sefusatti:2011cm, Anselmi:2011ef, D'Amico:2011pf, Ayaita:2011gp}.

The sound speed of dark energy can be defined as the relation between its pressure and density perturbations in the rest frame (corresponding to the comoving orthogonal gauge), $\hat{\dep}_X\equiv \cshat \hat{\der}_X$. It captures possible non-adiabatic pressure perturbations and therefore is, in general, not equal to the ratio $\bp_X'/\br_X'$. The rest-frame sound speed, $\hat{c_X}$, defines a relation between gauge-invariant combinations that can easily be transformed into any gauge \cite{Bardeen:1980kt,Kodama:1985bj,Hu:1998kj,Bean:2003fb}. Assuming for simplicity a constant equation of state we have
\be \label{genp}
\dep_X=\cshat\der_X+3(1+w_X)(\cshat-w_X)\frac{\ch}{k}\rho_XV_X
\ee
which is valid in any orthogonal gauge (and for any frame). 
We will focus on the case of small (positive) values of $\cshat$, which is the most interesting case observationally. Then, the first term on the right hand side of the expression above is negligible on large scales, $k\ll \H/\hat{c_X}$, and the dark energy component is approximately comoving with dark matter, $V_X\simeq V_m$, see e.g.\ \cite{Sapone:2010uy}. So we can write:
\be \label{genp1}
\dep_X\simeq-3w_X(1+w_X)\frac{\ch}{k}\rho_XV_m \,, \qquad \text{for}\ k\ll \H/\hat{c_X}
\ee
where the dark matter velocity can be read off from \eq{eq:Vm1} and \eq{eq:velevol}. Using the properties of the hypergeometric function 
$_2{F}_{1}$, this result can be approximated at $z=0$ by
\be \label{genp2}
\dep_X(z=0) \simeq-\frac{3}{4}w_X(1+w_X)\rho_X\left(1+\frac{1}{\Omega_m}\right)C_\Vk \,,
 \qquad \text{for}\ k\ll \H/\hat{c_X}
\ee
where $C_\Vk$ for $k\ll k_s$ is given in equation \eq{ck} in terms of the primordial power spectrum.

On small scales, deep inside the nonlinear regime (and hence below the sound horizon of dark energy), the relevant term of \eq{genp} is the one that depends on $\der_X$. In this limit it turns out that $\der_X$ is inversely proportional to $\cshat$ \cite{Sapone:2009mb}. Therefore the pressure fluctuation is again nearly independent of the sound speed
\be \label{applk}
\dep_X\simeq-(1+w_X)\rho_X\,\varphi \simeq -(1+w_X)\br_X \(\epsilon-\dfrac{5}{4}\alpha^{-3}\)\alpha^{-1}\,C_\Vk \,,  \qquad \text{for}\ k\gg \H/\hat{c_X}
\ee
where the Weyl potential $\varphi=C_\Vk {\cal F}(-\alpha^3)$ has been approximated by its late time expression \eq{limitphi} (where also $\epsilon\approx 1.44$ is defined), while $C_\Vk$ is given by the $k \gg k_s$ branch of \eq{ck}. These results show that the amplitude of the linear dark energy pressure perturbation does not depend on the value of the sound speed either for large or small scales. Naively, one would expect that nonlinear corrections would modify this feature on small scales. However, the results of \cite{Anselmi:2011ef} on the density power spectra indicate that the nonlinear corrections are roughly independent of the (small) value of the speed of sound.

What the sound speed does affect is the behaviour of the pressure perturbations at intermediate scales. This is shown in figure \ref{presscomp} where we plot the dimensionless power spectra, as defined in \eq{defpower}, of $\dep_m/(\br_m+\br_X)$ and $\dep_X/(\br_m+\br_X)$ at redshifts $0$ and $3$. Notice that we divide the pressure perturbation in each case by the total energy in the clustering DE model. Here we assume that $\dep_m$ computed in $\Lambda$CDM is roughly equal to $\dep_m$ in clustering wCDM. This is well justified as $\dep_m$ only depends on $V_m$ which is not affected strongly by dark energy clustering. The two models of dark energy depicted assume constant values of $w_X$ and $\cshat$. The amplitude at small and large scales is given by $w_X$ and is independent of $\cshat$. Hence, the equation of state controls the overall amplitude of the pressure perturbation and the sound speed determines the scale at which $\dep_X$ starts to drop off. 

The power spectrum of $\dep_m$ is computed at second order in the velocity perturbation. On large scales, the dark energy pressure perturbation is much larger than its counterpart from the frame effect on the matter. However, even though a complete nonlinear analysis is needed to study small scales ($k>0.1\,h/\Mpc$) the results indicate that for nonlinear scales the matter pressure can be comparable to the dark energy one. Therefore, for studies of dark energy perturbations on large scales using, for instance, the ISW effect on the CMB, the effective matter pressure can be safely ignored for interesting values of the dark energy parameters. On the other hand, for studies of dark energy clustering on smaller scales (e.g.\ WL), the frame effect will have to be taken into account for a reliable interpretation of the data. Such a study is beyond the scope of the present work and will involve an accurate computation of the dark energy perturbations on nonlinear scales \cite{D'Amico:2011pf, Anselmi:2011ef} including also the higher order contributions to \eq{deltape}.

\subsection{Anisotropic stress in modified gravity} \label{sec:megastress}

While the existence of a sound horizon and the associated pressure perturbations are a key feature of scalar field dark energy models like Quintessence and K-essence, the anisotropic stress is usually considered as a smoking gun for modifications of gravity \cite{Kunz:2006ca,Saltas:2010tt,Bertschinger:2011kk}. However, we have shown above that at nonlinear scales even $\Lambda$CDM exhibits a non-zero gravitational slip. This raises the following question: would it be possible to confuse this higher-order contribution with the anisotropic stress intrinsically present at the linear level in some MG models? Here we consider the DGP model which assumes that we live on a 4D brane embedded in a 5D bulk. Even though the self-accelerating solution has a ghost degree of freedom \cite{Luty:2003vm, Nicolis:2004qq} and, in addition, this model is observationally ruled out at the $5\sigma$ level (e.g.\ \cite{Davis:2007na, Lombriser:2009xg}), it is nonetheless one of the `standard' MG models as it exhibits many of the typical features of this class, and it has been well studied in the literature. Hence, for our purpose of comparing the matter anisotropic stress due to the frame effect to the anisotropic stress in MG, it will serve us well.

The gravitational DGP action is just the 5D Einstein-Hilbert action together with an induced 4D Einstein-Hilbert action confined to the brane. The relative strength of the two contributions is given by the crossover scale $r_c = M_P^2/M_5^3$ where $M_5$ is the 5D Planck mass. DGP therefore, like $\Lambda$CDM, is characterised by a single free parameter, the crossover radius $r_c$. The background expansion (assuming flatness) in DGP is given by (see e.g.\ \cite{Maartens:2006yt})
\be
H^2 - \frac{H}{r_c} = \frac{\br_m}{3M_P^2} \,
\ee
where $H$ is the cosmic time Hubble parameter. We will consider the second term on the left, $H/r_c$, as providing the dark energy contribution to the expansion rate and so think of it as being effectively on the right hand side. If the Hubble parameter today is given by $\H_0$ and the relative matter density today is taken to be $\Omega_m$, then we want for consistency $\H_0/r_c = \H_0^2 (1-\Omega_m)$ or $r_c = 1/[\H_0 (1-\Omega_m)]$. As naively expected, we see that the crossover scale of DGP needs to be of the size of the horizon scale today, which reveals that the fine-tuning required in this model is comparable to the one in $\Lambda$CDM. In the quasi-static limit relevant for scales that are subhorizon but still linear (see e.g.\ \cite{Koyama:2005kd}), one finds for the perturbations that
\bea
-k^2 \phi_1 &=& \frac{a^2\br_m}{2M_P^2} \(1 - \frac{1}{3 \beta}\) \(\delta_m + \frac{3 \H}{k} V_m \) \label{eq:dgp_phi} \\
-k^2 \psi_1 &=& \frac{a^2\br_m}{2M_P^2} \(1 + \frac{1}{3 \beta}\) \(\delta_m + \frac{3 \H}{k} V_m\) \label{eq:dgp_psi}
\eea
for $\beta \equiv 1-2 r_c H [1+\dot{H}/(3 H^2)] = 1+ 2 r_c H w_{X}$, and the matter perturbations evolve as usual as a pressure-free fluid sourced by the gravitational potentials above. However, since the Poisson equations for the potentials are altered, also the matter evolution is now different. Defining $\eta_1$, the MG parameter related to the anisotropic stress, as in \cite{Amendola:2007rr} we find that
\be
\eta_1 \equiv - \frac{\Pi_1}{\phi_1} = \frac{2}{3\beta -1} \, .  \label{eq:dgp_eta}
\ee
Numerically, $\eta_1$ is small at high redshifts but tends towards $\eta_1 \approx -0.44$ today for a flat Universe with $\Omega_m = 0.3$. In other words, the effective anisotropic stress in DGP on linear scales is of comparable size as the gravitational potentials themselves, and much bigger than the second order anisotropic stress in $\Lambda$CDM. Current constraints on the MG parameters are very weak, with deviations of order unity still allowed even for fairly restrictive parameterisations (see e.g.~\cite{Bean:2010zq,Daniel:2010yt,Song:2010fg,Zhao:2011te} for recent analyses).

Once the matter perturbations become nonlinear, the situation is more complicated. All viable MG models, specifically all models that pass solar system tests, need to screen the extra degrees of freedom at small scales since they would induce a fifth force of a strength far beyond observational limits. This screening effect is usually coupled to the local density through a Vainshtein \cite{Vainshtein:1972sx}, Chameleon \cite{Khoury:2003aq} or similar mechanism (e.g.\ \cite{Pietroni:2005pv}) and thus becomes active on small scales. In the case of DGP it is the Vainshtein mechanism that provides the necessary screening, and it is generically active on scales of the order of \cite{Lue:2002sw,Lue:2004rj}
\be
r_* = \left( \frac{r_c^2 r_S}{\beta^2} \right)^{1/3}  \simeq \left( \frac{r_S}{\H_0^2} \right)^{1/3} 
\ee
where $r_S = 2 G  M$ is the Schwarzschild radius of the source (here assumed to be a point mass). For a cluster with $M=10^{15} M_\odot$ the Vainshtein radius is of the order of $r_* \approx 10$ Mpc, which indeed coincides roughly with the scales at which perturbations start to become nonlinear. On smaller scales, ref.~\cite{Koyama:2007ih} estimates the anisotropic stress to be
\be
\Pi_1 \simeq \left( \frac{H^2 a^4}{r_c^2 k^4} \right) \delta_m
\ee
which means it is strongly suppressed on small scales, and roughly goes as $-(r_c k)^{-2}\phi_1$.

To summarise the discussion, the power spectrum of $\Pi_1$ in the DGP scenario has the following order of magnitude today:
\be
k^3 P_\Pi(k) \simeq  k^3 P_m(k) \times \left\{  
\begin{array}{cc} (\H_0/k)^2 & \qquad k \lesssim 1/r_*
\vspace{1mm} \\  (\H_0/k)^4 & \qquad k \gtrsim 1/r_* \, . \end{array} \right.
\ee
The anisotropic stress is therefore large on scales above roughly $0.1/$Mpc and decays more rapidly on smaller scales. The exact behaviour would have to be studied with numerical simulations. Comparing the small size of the matter anisotropic stress relative to the Weyl potential shown in figure \ref{fig:piby2phi} to the order unity effect in DGP shows that on linear and weakly nonlinear scales the MG signal is not contaminated by the matter velocity induced gravitational slip. The difference is so large that we expect the dynamical anisotropic stress in MG models to remain smaller than the effective matter anisotropic stress on all scales of interest for future large-scale surveys like e.g.\ Euclid\footnote{Euclid consortium webpage: http://www.euclid-ec.org} \cite{EditorialTeam:2011mu}.

\subsection{Confusion with galaxy bias?} \label{sec:biasonbias}

Finally, there is the question of whether it is possible to construct combinations of observables that are sensitive to the effective anisotropic stress and pressure. And, if the answer is positive, can we disentangle those signals from other uncertainties? A good example is the reconstruction of the galaxy bias by cross-correlating galaxy number counts and WL observables \cite{Pen:2004rm, Song:2008xd, Gaztanaga:2011yi}.

For illustration, let us briefly discuss this in linear theory with a slight gravitational slip, $\Pi\equiv\phi-\psi\neq 0$. Our aim here is to clarify how the presence of a gravitational slip, if ignored, would be misinterpreted as a non-trivial galaxy and velocity bias. Well below the Hubble scale, the observed galaxy number density fluctuation in the direction $\Vnhat$ is
\be
\Delta(\Vnhat,z) = \delta_g - \mu_k \frac{k}{\H}V_g -2\kappa
\ee
in terms of the actual galaxy number density fluctuation in the longitudinal gauge, $\delta_g$, the galaxy peculiar velocities, $V_g$, and their WL convergence, $\kappa$ (see for instance \cite{Yoo:2010ni,Challinor:2011bk,Bonvin:2011bg,Jeong:2011as})\footnote{Here we neglect the corrections from the Doppler term, the ISW effect, the time delay and the metric potentials at the galaxy positions for simplicity and also because they are negligible well inside the Hubble horizon.}. Here $\mu_k\equiv\Vnhat\!\cdot\!\hat\Vk$. We apply a simple local model for the Lagrangian galaxy bias where the number density, $g$, is a function of the comoving (rest frame) matter density, $\rho_m^*$, and the proper time, $\tau$. In the longitudinal gauge (normal frame) we can then write 
\be
\delta_g = b\delta_m +3b\frac{\H}{k}V_m +\frac{b_\tau}{k}V_g \,,\qquad
V_g = b_v V_m
\label{eq:biasdef}
\ee
where we define the Lagrangian galaxy bias $b$ and the evolution bias $b_\tau$
\be 
b \equiv \left(\frac{\partial \bar g}{\partial\rho_m^*}\right)_\tau \quad\,,\quad b_\tau \equiv \left(\frac{\partial\ln\bar g}{\partial\tau}\right)_{\rho}
\ee
as derivatives of the average galaxy number density $\bar g$. The velocity bias $b_v$ is often taken to be unity but we keep it here for clarity.

We now write all the perturbations in terms of the metric potentials. On subhorizon scales, and neglecting time derivatives of the potentials, we have from (\ref{eq:derG1}) and (\ref{eq:deqG1}) that
\be
\delta_m &\simeq& -\frac{2}{3\Omega_m(z)}\frac{k^2}{\H^2}\(\vf +\frac{1}{2}\Pi\)
\label{eq:dmBias}
\\
V_m &\simeq& \frac{2}{3\Omega_m(z)} \frac{k}{\H}\(\vf -\frac{1}{2}\Pi\) \,.
\label{eq:vmBias}
\ee
The WL convergence of sources at conformal distance $\chi_s$ is
\be
\kappa(\Vnhat,\chi_s) = \nabla_{\Vnhat}^2\int_{0}^{\chi_s}d\chi\, W(\chi_s,\chi)\, \vf(\chi_s\Vnhat,\eta_0-\chi)
\ee
where $\nabla_{\Vnhat}^2$ is the Laplacian on the unit sphere and $\eta_0$ is the conformal time today. The WL efficiency function in a flat background is $W(\chi_s,\chi)\equiv (\chi_s-\chi)/(\chi_s\chi)$. Its effect is a very broad 3D to 2D projection while weighing down contributions of lenses close to the sources as compared to those close to the observer. Therefore it is convenient to cross-correlate the galaxy number counts at a given redshift, $\Delta_g=\Delta(\Vnhat,z_g)$, with the WL signal of the background sources at higher redshift, $\kappa_b=\kappa(\Vnhat,z_b)$, such that the galaxies select a thin slice from the broad WL efficiency function. Well inside the horizon we can safely neglect the velocity term $\propto\H/k$ in (\ref{eq:biasdef}) such that
\be
\la\Delta_g\,\kappa_b\ra \simeq \left\la b\delta_m\,\kappa_b \right\ra
-\H^{-1} \la b_v \mu_k k V_m\,\kappa_b\ra -2\la\kappa\,\kappa_b \ra \,.
\ee
Using the expressions for $\delta_m$ and $V_m$ above, we can write schematically
\be
\frac{\la\Delta_g\,\kappa_b\ra}{2\la\kappa\,\kappa_b\ra} + 1 =
-\frac{\la k^2[ b(1+\Pi/2\vf) + b_v(1-\Pi/2\vf)]_g \vf_g\, W_b \vf_b \ra} 
{3\H^2(z_g)\Omega_m(z_g)\,\la W_g\vf_g\, W_b\vf_b \ra} 
\ee
where the line-of-sight integrations of the WL convergence are implicitly assumed. It is now easy to see that the presence of anisotropic stress, or a gravitational slip, leads to replacing the standard biases by
\be
b\to b\(1+\frac{\Pi}{2\vf}\) \,,\qquad b_v \to b_v\(1-\frac{\Pi}{2\vf}\) \,.
\label{eq:modBias}
\ee
Or the other way round: ignoring the presence of a gravitational slip biases the measurement of galaxy biases as shown above. Using different tracers of the density field at different redshifts one can now disentangle the biases and the gravitational slip. The corrections from second order perturbations are very small, see figure \ref{fig:piby2phi}. However, this result holds for any metric theory of gravity and most MG models do generate significant anisotropic stress as discussed in the previous section. Therefore it is in general important to take this effect on the bias into account.

In~\cite{Song:2008xd}, an elegant non-parametric approach was developed to test GR on cosmological scales using a set of consistency relations involving metric and matter density perturbations in linear theory. It was pointed out that to test any of these relations, the observations of peculiar velocities and galaxy clustering can be combined to predict the outcome of WL measurements in order to have a strong self-consistency test of $\Lambda$CDM and measure the linear galaxy bias similarly as discussed here. While this proposal is consistent at linear order, we would like to point out that, in any case, nonlinearities induce anisotropic stress that needs to be taken into account properly when undertaking such a program with information on nonlinear scales. Furthermore, as we see above, if a gravitational slip is actually present in linear theory in reality (as in MG), it is readily mistaken for a non-trivial galaxy bias if not taken into account.

\section{Conclusions}  \label{sec:conclusions}

With current cosmological observations tightening constraints on the $\Lambda$CDM scenario and not (yet) hinting at strong deviations from it, or from GR, it becomes increasingly important to devise ways of falsifying or confirming the standard paradigm. To be able to do so, a very precise understanding of the observations and the corresponding predictions is needed not only on linear but also on nonlinear scales. In recent years, much progress has been made in nonlinear Newtonian perturbation theory to understand and complement numerical analyses. Simulation efforts in non-standard scenarios are also underway. On the other hand, relativistic perturbation theory predictions and consistency relations are mostly restricted to linear theory. But to be able to confidently falsify or confirm the standard paradigm it is imperative to understand the implications of dynamics in curved space on weakly nonlinear scales where Newtonian theory may not be sufficient, even more so if modifications to GR are allowed.

A framework based on casting beyond-$\Lambda$CDM structure growth in terms of two functions, that can either be parameterised in generic ways or predicted from a given theory, has recently received considerable attention. This tool is often referred to as Parameterised-Post-Friedmann approach or modified growth parameterisations. In this paper, we propose an alternative framework deriving fully nonlinear consistency relations between the geometry, the matter content and any possible dark component. Since our framework requires neither the assumption of GR nor minimal interactions, but only relies on the symmetries of the Einstein tensor, it can be applied to different scenarios and models. The framework derives from the covariant approach, employing a very practical choice of 4-vector field (the normal frame) to decompose the Einstein tensor, the energy-momentum tensor and their difference, $X^\mn\equiv M_P^2 G^\mn - T^\mn$, that describes the physics beyond GR and/or beyond the standard constituents of the cosmological model. The normal frame, $n^\mu$, is defined to be orthogonal to the surfaces of constant time. Because its definition is purely geometric it is only given by the metric itself and so are the irreducible components of $M_P^2G^\mn$: the \emph{geometrical} dynamic quantities. This is the reason why the normal frame is such a convenient choice, as it naturally allows us to write consistency relations that separate into a geometrical part, a matter part and a dark energy part:
\be \nn
p_X = p_G - \frac{1}{3}v_m^2\rho_m \,, \qquad
\pi_X^\mn = \pi_G^\mn - \rho_m v_m^{\la\mu}v_m^{\nu\ra}
\ee
where the $G$-variables are only given by the metric, see section \ref{sec:fullconsistency}. These fully nonlinear consistency relations can either be taken as conditions for $\Lambda$CDM+GR to hold by requiring $p_X=-M_P^2 \Lambda$, $\pi_X^\mn=0$, or the $X$-quantities can be parameterised and reconstructed from data. The advantage of this very general framework is that it is a top-down approach that a priori takes all non-linearities and all scalar, vector and tensor degrees of freedom into account, and can be simplified in perturbation theory at will.

As a consequence of selecting the normal frame, the dynamical quantities on the matter side (the energy density, the pressure, the energy flux and the anisotropic stress) do not take the values that one would expect, or find when measured in the matter rest frame. The transformations from the rest frame to the normal frame can be written fully nonlinearly. These show that a non-vanishing effective pressure and anisotropic stress are measured in the normal frame, even if the matter is assumed to be a perfect pressureless fluid in its rest frame, see section \ref{sec:framechange}. This is the origin of the terms $\sim v_m^2$ in the consistency relations. However, these effects are only relevant at second order in perturbation theory and they can be accounted for, knowing that they are present. Furthermore, we point out that cosmological observations do not actually measure the density field in its comoving frame, as we do observe the effect of the velocity field of the sources through the projection onto the photon wave vector. This means that  there are non-comoving effects that need to be taken into account. More importantly, apart from gravitational nonlinearities, we identify the pressure and anisotropic stress measured in the normal frame as the physical quantities that enter as sources in the Bardeen equation for the second order Weyl potential and the constraint for the second order gravitational slip. This is a consequence of the fact that the normal frame is purely defined in terms of the metric. Finally, let us recall that the normal frame helps to recover the Eulerian picture of Newtonian perturbation theory from GR. Although GR is a theory of curved spacetime, the normal frame is related to the Eulerian picture in Newtonian nonlinear dynamics.

We compute the consistency relations for $\Lambda$CDM in first and second order scalar perturbation theory in the generalised longitudinal gauge (that is the gauge where $n^\mu$ has zero spatial components and does not see shear or vorticity). At first order we recover the usual conditions: the gravitational slip vanishes, $\Pi_1\equiv\phi_1-\psi_1=0$, and the Weyl potential, $\vf_1\equiv\frac{1}{2}(\phi_1+\psi_1)$, follows the standard Bardeen equation, see section \ref{sec:linear}. We solve the Bardeen equation analytically to provide the basis for a careful study of the matter pressure and anisotropic stress. We compute $\frac{1}{3}v_m^2\rho_m$ and $\rho_m v_m^{\la\mu}v_m^{\nu\ra}$ at second order and find these terms to be small on large scales. They become relevant only on scales $k\gtrsim 0.1\, h/\Mpc$. The second order anisotropic stress consistency relation reveals a non-vanishing gravitational slip due to the effective matter anisotropic stress and non trivial combinations of products of spatial derivatives of $\vf_1$, see equation \eqref{eq:k4pi} and also refs. \cite{Acquaviva:2002ud,Vernizzi:2004nc}. Here we clarify the origin of the different contributions to $\Pi_2$. The pressure consistency, on the other hand, yields the second order analogue of the Bardeen equation for $\vf_2$, see equation \eqref{eq:bardeen2}. It acquires different source terms which we analyse, compute and compare and we find subtle cancellations between the contributions from the effective matter stresses and $\partial^2\vf_1^2$--type terms. The fact that the source term of the second order Bardeen equation depends non trivially and sensitively on the pressure and anisotropic stress perturbations of a possible dark energy makes the weakly non-linear evolution a promising tool to test the $\Lambda$CDM paradigm. Combining different cosmological probes enables us to reconstruct the consistency relations, as all observations probe the metric potentials, either directly or indirectly via the Poisson equation, see also e.g.\ \cite{Song:2008xd}. Let us emphasise that the effective matter pressure and anisotropic stress are seen by the second order metric perturbations in any case, irrespectively of the yet to be solved issue of which precise frame is the one that corresponds to us as observers. 

Finally, we discuss a set of applications: we compare the effective matter pressure with the pressure perturbations in clustering dark energy, concluding that the matter pressure is negligible on large scales and becomes relevant on weakly nonlinear scales. Thus, the matter pressure only needs to be taken into account when deriving constraints on the dark energy sound speed from small scale observations. Furthermore, we compare the matter anisotropic stress to typical anisotropic stresses arising in MG scenarios, specifically the DGP model. Anisotropic stress is normally considered a smoking gun for the detection of MG. We find that, typically, the matter anisotropic stress is by far subdominant and not an issue. Finally, we assess the impact of the anisotropic stress on the determination of galaxy bias from combining WL and galaxy clustering data. We find that the gravitational slip affects the Lagrangian galaxy bias and the velocity bias in opposite ways, but only marginally. Using multiple tracers at several redshifts and different combinations of observables, we expect that one could be able to filter out or strongly constrain the gravitational slip.

\acknowledgments

We thank Vincent Desjacques, Ruth Durrer, Roy Maartens, Sabino Matarrese and Massimo Pietroni for helpful discussions. Guillermo Ballesteros is supported by the Centro Enrico Fermi and INFN; and thanks the D\'epartement de Physique Th\'eorique of the Universit\'e de Gen\`eve for its hospitality. This work has been supported by the Swiss National Science Foundation.

\appendix

\section{Conventions}  \label{app:conventions}

We employ natural units where $c=\hbar=k_B=1$ and define the reduced Planck mass as $M_P=(8\pi G)^{-1/2}$, with $G$ being Newton's constant as measured in laboratory experiments. We work with a signature \mbox{($-$ + + +)}. For tensor components, Greek indices take values $0\ldots3$, while Latin indices run from $1$ to $3$. Round parenthesis around indices mean symmetrisation, $Y^{(\mn)}\equiv \frac{1}{2}(Y^\mn+Y^{\nu\mu})$, and square brackets mean anti-symmetrisation, $Y^{[\mn]}\equiv \frac{1}{2}(Y^\mn-Y^{\nu\mu})$.

The spatially flat, homogeneous and isotropic FLRW metric is $\bar g_\mn dx^\mu dx^\nu \equiv a^2(-d\eta^2 +\delta_{ij}dx^idx^j)$, where $\eta$ is the conformal time and $a(\eta)$ is the scale factor normalised to unity today so that the comoving scales are numerically equal to physical scales, at present time. Derivatives with respect to $\eta$ are denoted with a prime and the conformal Hubble parameter is $\H\equiv a'/a = aH$ where $H$ is the cosmic time Hubble parameter, whose value today is the Hubble constant $\H_0=100h\, \km/{\rm s}/\Mpc\approx 3\times 10^{-4}\,h/\Mpc$. Overbars denote background quantities and subindices in perturbation variables indicate the order of the fluctuation. In our notation, $\Omega_i$ represents the present relative energy density of the $i$--component, and is therefore a constant unless specified otherwise.

When numerically computing results for a typical flat $\Lambda$CDM model we use the following parameter values: $A_s= 2.1\times 10^{-9}$, $n_s = 0.96$, $k_p = 0.05\,/\Mpc$, $h = 0.7$, $\Omega_c h^2 = 0.112$, $\Omega_b h^2 = 0.0226$ and $\Omega_rh^2 = 4.17\times 10^{-5}$.

Our convention for the Fourier transform is
\be
f(\Vk) = \int d^3x\, f(\Vx) e^{i\Vk\cdot\Vx}\,,
\qquad \text{and hence}\qquad
f(\Vx) = \int \frac{d^3k}{(2\pi)^3}\, f(\Vk)\, e^{-i\Vk\cdot\Vx} \,.
\ee
The power spectrum of any random variable $f$ is given by the Fourier transform of the two-point correlation function:
\be
(2\pi)^3 \delta_D(\Vk-\Vk') P_f(k) \equiv \la f(\Vk) f^*(\Vk') \ra 
\ee
where $\delta_D(\Vk-\Vk')$ denotes the 3D Dirac delta distribution. Finally,
\be \label{defpower}
{\cal P}_f(k) \equiv \frac{k^3}{2\pi^2} P_f(k)
\ee
defines the dimensionless power spectrum.

\section{Covariant 1+3 decomposition}  \label{app:covariant}

This appendix gives a brief summary and an account for our notation of the covariant approach to cosmology. More details on the formalism and its applications can be found for example in \cite{Ellis:1971pg, Ellis:1989jt, Bruni:1992dg, Ehlers:1993gf, Maartens:1998xg,Clarkson:2010uz}.

\subsection{Defining the covariant formalism}  \label{app:covariant_defs}

The covariant approach defines a 1+3 splitting of space-time by using a time-like unit 4-velocity field, $u^\mu$ with $u^2\equiv g_\mn u^\mu u^\nu = -1$. It represents a family of observers and is often referred to as an observer frame. All tensors are decomposed into their irreducibles w.r.t.\  $u^\mu$, the projection tensor $\P^\mn\equiv g^\mn+u^\mu u^\nu$ and the projected alternating tensor $\veps_{\mn\alpha}\equiv\eta_{\mn\ab}u^\beta$ where $\eta_{\mn\ab}\equiv -\sqrt{-g}\delta^0_{[\mu}\delta^1_\nu \delta^2_\alpha \delta^3_{\beta]}$. Then a dot and $D_\mu$ represent the covariant derivative parallel and orthogonal to $u^\mu$, respectively, which for a generic tensor $J$ read
\bea
\dot J^{\mu\cdots}_{\phantom{\mu\cdots}\cdots\nu} &\equiv&
  u^\alpha\nabla_\alpha J^{\mu\cdots}_{\phantom{\mu\cdots}\cdots\nu}
\\
D_\lambda J^{\mu\cdots}_{\phantom{\mu\cdots}\cdots\nu} &\equiv&
  \P_\lambda^\alpha \P^\mu_\beta \cdots \P_\nu^\gamma \nabla_\alpha
  J^{\beta\cdots}_{\phantom{\beta\cdots}\cdots\gamma}
\eea
To describe the geometry as seen by the family of observers $u^\mu$, one uses the irreducible components of $\nabla_\nu u_\mu$
\be
\nabla_\nu u_\mu = -\dot u_\mu u_\nu +\frac{1}{3}\Theta\P_\mn
  +\veps_{\mn\alpha}\omega^\alpha +\sigma_\mn \,.
\ee
These are the \emph{kinematic or geometric quantities} of $u^\mu$: the expansion $\Theta\equiv D^\alpha u_\alpha$, the acceleration $\dot u_\mu \equiv u^\alpha\nabla_\alpha u_\mu$, the shear $\sigma_\mn\equiv D_{\la\mu}u_{\nu\ra}$, and the vorticity $\omega_\mu\equiv-\frac{1}{2}\veps_{\mu\ab}D^\alpha u^\beta$. Here the projected symmetric trace-free parts are defined as
\be
V^{\la\mu\ra} \equiv \P^\mu_\alpha V^\alpha \,,\quad
Y^{\la\mu\ra\nu} \equiv \P^\mu_\alpha Y^{\alpha\nu} \,,\qquad
Y^{\la\mn\ra} \equiv \[\P^{(\mu}_{\alpha\phantom{\beta}} \P^{\nu)}_\beta
  -\frac{1}{3}\P^\mn\P_{\ab}\] Y^{\ab} \,.
\ee

In the covariant approach we decompose $Y^\mn\in\{M_P^2 G^\mn,\, T^\mn,\, X^\mn\}$ into its \emph{dynamic quantities} or \emph{fluid variables}: energy density $\rho_Y\equiv u_\alpha u_\beta Y^\ab$, isotropic pressure (including possible bulk viscous stress) $p_Y\equiv\frac{1}{3}\P_\ab Y^\ab$, energy flux (heat plus particle flux) $q_Y^\mu\equiv -u_\alpha Y^{\la\mu\ra\alpha}$ and anisotropic stress (shear viscous stress) $\pi_Y^\mn\equiv Y^{\la\mn\ra}$. Then, we can write
\be
Y^\mn = \rho_Y u^\mu u^\nu + p_Y \P^\mn + 2q^{(\mu}_Y u^{\nu)} + \pi^\mn_Y
\ee
and the dynamic quantities add up as follows, see \eqref{eq:Xdef}, 
\be
\rho_G = \rho_T + \rho_X \,,\qquad
p_G = p_T + p_X \,,\qquad
q^\mu_G = q^\mu_T + q^\mu_X \,,\qquad
\pi^\mn_G = \pi^\mn_T + \pi^\mn_X
\label{eq:totalvariables}
\ee
where the subscript $G$ indicates the ``geometrical'' fluid variables defined by projecting $M_P^2G^\mn$. Notice that this is still fully nonlinear and generic, and these are dynamical and constraint equations for the metric and the possible additional dark degrees of freedom, but there are no derivatives of the matter variables coming in. Usually we see the $G$-variables simply as the total energy density, pressure, etc.

\subsection{Change of frame}  \label{app:framechange}

The fact that each component has an energy flux is because we decompose the multi-component system w.r.t.\  a generic observer frame $u^\mu$ rather than a specific 4-velocity of one of the components or the total energy-momentum content. However, the physical properties such as the sound speed or the barotropic equation of state of a given component are described in its energy-frame, which means in terms of the dynamic quantities as observed by its 4-velocity $u_Y^\mu$, denoted by an asterisk. The energy-frame is the frame in which the energy-flux vanishes, $q_Y^{*\mu}\equiv 0$ and in most cases coincides with the particle rest frame. The energy-frame satisfies the eigenvalue equation $Y^\mu_\alpha u_Y^\alpha =-\rho^*_Y u_Y^\mu$. In terms of the energy-frame dynamic quantities the energy-momentum tensor reads
\be
Y^\mn = \rho^*_Y u_Y^\mu u_Y^\nu + p^*_Y \P_Y^\mn + \pi^{*\mn}_Y
\ee
where $\P_Y^\mn\equiv g^\mn + u_Y^\mu u_Y^\nu$. The transformation from the rest frame $u_Y^\mu$ to a generic frame $u^\mu$ depends on the relative velocity of the two frames, $v_Y^\mu$. It is given through the relation $u_Y^\mu = \gamma_Y(u^\mu + v_Y^\mu)$ that is derived by decomposing $u_Y^\mu$ w.r.t.\ $u^\mu$ and using the normalisation conditions $u_Y^2=-1=u^2$. Therefore $v_Y^\mu u_\mu = 0$ and $\gamma_Y=(1-v_Y^2)^{-1/2}$. The dynamic quantities in the rest frame transform to the generic $u^\mu$-frame as
\be
\rho_Y &=& \rho_Y^* + \[\gamma_Y^2 v_Y^2(\rho_Y^*+p_Y^*)
  +\pi_Y^{*\ab}v_{Y\alpha}v_{Y\beta}\]
\\
p_Y &=& p_Y^* + \frac{1}{3}\[\gamma_Y^2 v_Y^2(\rho_Y^*+p_Y^*)
  +\pi_Y^{*\ab}v_{Y\alpha}v_{Y\beta}\]
\\
q_Y^\mu &=& (\rho_Y^*+p_Y^*)v_Y^\mu + \[\gamma_Y^2v_Y^2(\rho_Y^*+p_Y^*)v_Y^\mu
  +\pi_Y^{*\mu\alpha}v_{Y\alpha} - \pi_Y^{*\ab}v_{Y\alpha}v_{Y\beta}u^\mu\]
\\
\pi_Y^\mn &=& \pi_Y^{*\mn} +\[ \gamma_Y^2(\rho_Y^*+p_Y^*)v_Y^{\la\mu}v_Y^{\nu\ra}
\right. \nn \\ && \left.
  +\pi_Y^{*\alpha\beta}v_{Y\alpha}v_{Y\beta}\Big(u^\mu u^\nu -\frac{1}{3}\P^\mn\Big)
  -2u^{(\mu}\pi_Y^{*\nu)\alpha}v_{Y\alpha} \]
\ee
see for instance the appendix of ref.\ \cite{Clarkson:2010uz} specialised to $u_Y^\mu$ being the rest frame of the $Y$-component, i.e.\ $q_Y^{*\mu}=0$. The terms in square brackets are at least second order in the relative velocity and only the energy flux is non trivial at first order.

\subsection{Conservation equations}  \label{sec:conserv}

The connection between kinematic and the dynamic quantities is made via $\nabla_\alpha G^{\mu\alpha}=0$ a consequence of the Bianchi identities and therefore a geometrical fact (or total energy-momentum conservation if you wish):
\be
\hspace{-5mm}
\dot\rho_G + \Theta(\rho_G+p_G) + \(D_\alpha + 2\dot u_\alpha\) q_G^\alpha
  + \sigma_\ab\pi^{\ab}_G &=& 0
\\
\hspace{-5mm}
\dot q^G_{\la\mu\ra} +\frac{4}{3}\Theta q^G_\mu + \dot u_\mu(\rho_G+p_G) + D_\mu p_G
  + \(\sigma_{\mu\alpha} + \veps_{\mu\ab}\omega^\beta \) q_G^\alpha
  + \(D^\alpha + \dot u^\alpha\) \pi^G_{\mu\alpha} &=& 0
\ee
which are the continuity and the Euler equations, respectively, and are not independent of the field equations above. Note that the total $G$-variables can only be replaced by the $T$- or $X$-variables under the assumption that the respective energy-momentum tensors are separately conserved.

\section{Metric perturbations and the normal frame}  \label{app:metric}

\subsection{The kinematic quantities of the normal frame in the longitudinal gauge}  \label{app:kinematic}

The normal frame $n^\mu$ defines a set of kinematic quantities in the covariant approach: acceleration $\dot n_\mu$, the isotropic expansion $\Theta$, the shear $\sigma_\mn$ and the vorticity $\omega_\mu$. Up to second order in cosmological perturbations in the generalised longitudinal gauge they read
\be
\dot n^\mu &=& \left\{ \partial_i\Big[ \vf -\frac{1}{2}\Pi \Big]
  +\Pi_1\partial_i\Big[ 2\vf_1 -\Pi_1 \Big] \right\} \,a^{-2}\delta^{\mu i}\\
\dot n_\mu &=& \partial_i \[ \vf -\frac{1}{2}\Pi
-\Big(\vf_1^2-\frac{1}{2}\Pi_1\Big)^2 \] \, \delta^i_\mu
\\
\Theta(n) &=& 3H\left\{ 1 -\vf -\frac{\vf'}{\H} +\frac{1}{2}\Pi -\frac{\Pi'}{2\H}
  +\frac{3}{2}\Big(\vf_1-\frac{1}{2}\Pi_1\Big)^2 
\right. \nn \\  && \left.
  -\H^{-1}\Big(\vf_1+\frac{3}{2}\Pi_1\Big)\Big(\vf_1+\frac{1}{2}\Pi_1\Big)' \right\}
\\
\sigma_\mn(n) &=& 0
\\
\omega_\mu(n) &=& 0 \,.
\ee
Their covariant definitions are given in appendix \ref{app:covariant_defs}.

\subsection{Exact solutions of perturbation evolution after radiation-matter equality}  \label{app:pertevol}

The Bardeen equation after radiation-matter equality \eqref{eq:lin_cond} can be solved analytically in $\Lambda$CDM where $w_G=-\Omega_\Lambda/(\Omega_\Lambda+\Omega_m a^{-3})$. We find
\be \label{eq:lin_phi_full}
\vf_1 = C_\Vk {\cal F}\!\(-\alpha^3\) + D_\Vk\sqrt{\alpha^{-5} + \alpha^{-2}}
\ee
where $\alpha\equiv (\Omega_\Lambda/\Omega_m)^{1/3}a$ is the scale factor normalised at $\Lambda$-matter equality. The second mode, $\propto D_\Vk$, is decaying rapidly and can safely be ignored. The time-evolution of the first mode, $\propto C_\Vk$, is given by the ordinary hypergeometric function ${\cal F}(y)\equiv {}_2F_1(1/3,\,1;\, 11/6;\, y)$. To see its early and late time behaviour we expand the solution for small and large $\alpha$ to see the limiting behaviour:
\be \label{limitphi}
{\cal F}\!\(-\alpha^3\) \simeq \left\{ \begin{array}{ll}
 \(1-\dfrac{2}{11}\alpha^3\) \qquad &{\rm for}\ \ \alpha\ll 1
\vspace{2mm}\\
 \(\epsilon-\dfrac{5}{4}\alpha^{-3}\) \alpha^{-1} \qquad &{\rm for}\ \ \alpha\gg 1
\end{array} \right.
\label{eq:lin_phi_approx}
\ee
where $\epsilon\equiv 2\pi^{-1/2}\Gamma(2/3)\Gamma(11/6)\approx 1.44$. The early-time approximation is accurate to $<1\%$ for $a<0.5$ and to $<10\%$ for $a<0.8$, while at $a=1$ the late time approximation is more accurate than the early-time approximation, although only good to about $22\%$. 

We derive the matter density contrast and the velocity perturbation from the exact solution for $\vf_1$ by means of (\ref{eq:derG1}) and (\ref{eq:deqG1}):
\be
\delta_{m1} &=& \frac{2 a\,\nabla^2 \vf_1}{3\H_0^2\Omega_m}
  - 2\(1+\alpha^3\)\(\vf_1 + \frac{\dd\vf_1}{\dln a}\)
\\
V_{m1}^i &=& -\frac{2}{3} (1+\alpha^3)\, \H^{-1}\partial_i
  \[ \vf_1 + \frac{\dd\vf_1}{\dln a} \]
\ee
where we used $\br_G/\br_m=(1+\alpha^3)$ in $\Lambda$CDM. Using the solution (\ref{eq:lin_phi_full}) with $D_\Vk=0$ we find for the matter density contrast
\be \label{matters}
\delta_{m1,\Vk} = -2 \( \frac{a {\cal F}(-\alpha^3)}{3\Omega_m} \, \frac{k^2}{\H_0^2} 
+ \(1+\alpha^3\) \[ {\cal F}\!\(-\alpha^3\)   -3\alpha^3\tilde{\cal F}\!\(-\alpha^{3}\)   \] \) C_\Vk
\ee
where $\tilde{\cal F}(y) \equiv (d/dy){\cal F}(y) = {}_2F_1(4/3, 2; 17/6; y)$. On subhorizon scales, $k\gg \H$, we recover the usual growing mode solution, $\delta_{m1}\propto a$, in the matter era, using the early-time behaviour of ${\cal F}$ given in (\ref{eq:lin_phi_approx}) for $\alpha\ll 1$. For the velocity, we find
\be
V_{m1,\Vk}^j &=& -i {\cal V}(a)\frac{k_j}{\H_0} C_\Vk
\\
{\cal V}(a) &\equiv& \frac{2}{3} 
  \[\frac{a(1+\alpha^3)}{\Omega_m}\]^{1/2} \[ {\cal F}\!\(-\alpha^3\) 
     -3\alpha^3 \tilde{\cal F}\!\(-\alpha^{3}\)     \]
\ee
which is also valid on all scales. At early times we have
\be
{\cal V} \simeq \frac{2}{3}\frac{a}{\Omega_m^{1/2}}
  \[ 1 -\frac{8}{11}\alpha^3 +\frac{112}{187}\alpha^6 \]
  \,,\qquad &{\rm for}\ \ \alpha\ll 1
\ee
an approximation that is accurate to $<1\%$ for $a<0.4$ and to $<10\%$ for $a<0.6$.

\subsection{Second order geometrical fluid variables after radiation-matter equality}
\label{app:secondordergeometrical}

We compute the geometrical fluid variables, $\der_{G2}$, $\dep_{G2}$, $\deq_{G2}^\mu$ and $\depi_{G2}^\mn$ at second order in the generalised longitudinal gauge. These are the irreducible dynamic variables projected from $M_P^2G_\mn$, see section \eqref{app:covariant_defs}. To simplify the expressions, we already impose the $\Lambda$CDM condition for the first order gravitational slip, $\Pi_1=0$. We find
\be
\der_{G2} &=& -2\bar{\rho}_G \left\{ \vf_2 + \frac{\vf_2'}{\H} - \frac{\nabla^2 \vf_2}{3\H^2}
  -\frac{1}{2} \Pi_2 + \frac{\Pi_2'}{2 \H} - \frac{\nabla^2 \Pi_2}{6 \H^2}
\right. \nn \\ && \left. \hspace{2cm}
  -4\vf_1^2 -8\vf_1 \frac{\nabla^2 \vf_1}{3\H^2} -\frac{(\nabla\vf_1)^2 +(\vf_1')^2}{\H^2} \right\}
\\
\label{eq:depG2}
\dep_{G2} &=& \frac{2}{3} \bar{\rho}_G \left\{  -3w_G \vf_2 + 3\frac{\vf_2'}{\H} + \frac{\vf_2''}{\H^2}
  +\frac{3}{2}w_G \Pi_2 + \frac{\Pi_2'}{2\H} + \frac{\Pi_2''}{2\H^2} - \frac{\nabla^2\Pi_2}{3\H^2}
\right. \nn \\ && \left. \hspace{2cm}
  + 12w_G \vf_1^2 - 8\vf_1 \frac{\vf_1'}{\H} - \frac{(\vf_1')^2}{\HH^2} - 8\vf_1 \frac{\nabla^2 \vf_1}{3\H^2} 
- \frac{7(\nabla \vf_1)^2}{3\H^2} \right\}
\\
\deq_{G2}^\mu &=& - \frac{2}{3} \br_G \H^{-1} \left\{ \partial_i \left[\vf_2 
  + \frac{\vf_2'}{\H} - \frac{1}{2}\Pi_2 + \frac{\Pi_2'}{2\H} \right]
\right. \nn \\ && \left. \hspace{2cm}
  + \partial_i \left[ 2\vf_1\frac{\vf_1'}{\H} - \vf_1^2 \right] 
  + 4\vf_1\frac{\partial_i\vf_1'}{\H} \right\}
  \, a^{-1} \delta^{\mu i}
\\
\label{eq:depiG2}
\depi_{G2}^\mn &=& \frac{1}{3} \br_G \H^{-2} \left\{ 
  \Big(\partial_i\partial_j -\frac{1}{3}\delta_{ij}\nabla^2\Big)\Pi_2
  +8\vf_1\Big(\partial_i\partial_j -\frac{1}{3}\delta_{ij}\nabla^2\Big)\vf_1
\right. \nn \\ && \left. \hspace{2cm}
  +4\Big( \partial_i\vf_1 \partial_j\vf_1 -\frac{1}{3}\delta_{ij}(\nabla\vf_1)^2 \Big) \right\}
   \, a^{-2} \delta^{i\mu}\delta^{j\nu} \,.
\ee
These expressions are used in section~\ref{sec:secconsistency} to derive the consistency relations at second order.

\bibliography{refsb}
\bibliographystyle{JHEP}

\end{document}